\documentclass[12pt,preprint]{aastex}
\usepackage[usenames]{color}

\def\l{$\lambda$}
\def\mbh{$M_{\rm BH}$\/}
\def\nh{$n_{\mathrm{H}}$\/}

\def\ne{$n_{\rm e}$\/}
\def\nc{$N_{\rm c}$\/}
\def\rfe{$R_{\rm FeII}$}

\def\feiiq{\rm Fe{\sc ii}$\lambda$4570\/}
\def\msol{M$_\odot$\/}

\def\ltsima{$\; \buildrel < \over \sim \;$}
\def\ltsim{\lower.5ex\hbox{\ltsima}}  
\def\gtsima{$\; \buildrel > \over \sim \;$}

\def\gtsim{\lower.5ex\hbox{\gtsima}}

\def\lya{{ Ly}$\alpha$}
\def\civ{{\sc{Civ}}$\lambda$1549\/}

\def\cmq{cm$^{-2}$\/}
\def\cm3{cm$^{-3}$\/}
\def\hb{{\sc{H}}$\beta$\/}

\def\hbbc{{\sc{H}}$\beta_{\rm BC}$\/}

\def\niv{{\sc{Niv]}}$\lambda$1486\/}
\def\ciii{{\sc{Ciii]}}$\lambda$1909\/}
\def\oiiiopt{{\sc{[Oiii]}}$\lambda\lambda$4959,5007\/}
\def\o4363{{\sc{[Oiii]}}$\lambda$4363\/}
\def\caii{{Ca{\sc ii}}}

\def\siiii{Si{\sc iii]}$\lambda$1892\/}
\def\aliii{Al{\sc iii}$\lambda$1860\/}
\def\heiiuv{He{\sc{ii}}$\lambda$1640}
\def\nv{{N\sc{v}}$\lambda$1240}

\def\feiiopt{{Fe \sc{ii}}$_{\rm opt}$\/}
\def\feii{{Fe\sc{ii}}\/}
\def\siii{{Si\sc{ii}}$\lambda$1814\/}
\def\feiii{{Fe\sc{iii}}\/}
\def\fe{{\sc{Fe}}\/}

\def\fe76087{{\sc [Fe vii]}$\lambda$6087\/}

\def\kms{km~s$^{-1}$}

\def\ergss{ergs s$^{-1}$\/}
\def\hii{H{\sc ii}\/}

\def\heii{{{\sc H}e{\sc ii}}$\lambda$4686\/}

\def\rb{$r_{\rm BLR}$\/}

\def\siiv{Si{\sc iv}$\lambda$1397\/}
\def\oiv{O{\sc iv]}$\lambda$1402\/}

\def\siiiuv{Si{\sc ii}$\lambda$1533\/}

\def\Gsoft{$\Gamma_{soft}$}

\slugcomment{}

\shorttitle{BLR conditions in extreme Pop. A quasars}
\shortauthors{Negrete et al.}

\begin{document}
\title{BLR Physical Conditions in Extreme Population A Quasars:\\  a Method to Estimate  Central Black Hole Mass at High Redshift}

\author{C. Alenka Negrete and Deborah Dultzin}
\affil{Instituto de Astonom\'ia, Universidad Nacional Aut\'onoma de M\'exico, Mexico}
\email{anegrete@astro.unam.mx, deborah@astro.unam.mx}

\author{Paola Marziani}
\affil{INAF, Astronomical Observatory of Padova, Italy}
\email{paola.marziani@oapd.inaf.it}

\and

\author{Jack W. Sulentic}
\affil{Instituto de Astrof\'isica de Andaluc\'ia, Spain}
\email{sulentic@iaa.es}

\begin{abstract}
We describe a method for  estimating physical conditions in the broad line region (BLR) for a significant subsample of Seyfert-1 nuclei and quasars. Several diagnostic ratios  based on  intermediate (\aliii, \siiii)  and   high (\civ, \siiv) ionization  lines in the UV spectra of quasars are used to constrain density, ionization and metallicity of the emitting gas. We apply the method to two extreme Population A quasars -- the prototypical NLSy1 I Zw 1 and higher $z$\ source SDSS J120144.36+011611.6. Under assumptions of spherical symmetry and pure photoionization we infer BLR physical conditions: low ionization (ionization parameter $< 10^{-2}$),  high density (10$^{12} - 10^{13}$ \cm3) and  significant metal enrichment. Ionization parameter and density can be derived independently for each source with an uncertainty that is  less than $\pm 0.3$ dex.  We use the product of density and ionization parameter  to estimate the BLR radius and derive an estimation of the virial black hole mass (M$_{BH}$).  Estimates of M$_{BH}$ based on the ``photoionization'' analysis described in this paper are probably  more accurate than those derived from  the mass -- luminosity correlations widely employed to compute black hole masses for high redshift quasars. 
\end{abstract}


\section{Introduction}

We  lack a simple diagnostic method to estimate physical conditions (density, ionization parameter, metallicity) in the broad line region (BLR) of quasars.  Techniques for estimating electron  density and ionization in nebular astrophysics \citep{osterbrockferland06} are not straightforwardly applicable to broad lines in quasars. Reasons include: large line widths, line doublets too closely spaced to resolve individual components, and density at least an order  of magnitude higher than the critical density assumed for forbidden transitions modeled in spectra of planetary nebul\ae\ and \hii\ regions. The ionization parameter $U$, 
that represents the dimensionless ratio of the number of ionizing photons and the electron density \ne\ or, equivalently, the total number density of hydrogen \nh\, ionized and neutral\footnote{ In a fully ionized medium \ne\ $\approx 1.2$ \nh. We prefer to adopt a definition based on \nh\ because it is the one employed in CLOUDY computations.}, can be estimated from the intensity ratio of two strong  resonance lines arising from different  ionic stages of the same element \citep{davidsonnetzer79}.  It is again not easy to interpret the results in the case of quasars. For example, the ratio  \ciii/\civ\ may not yield a meaningful value if the relatively  low \ciii\  critical density  implies that the line is  formed at larger radii than \civ.  Using lines of different ions widens the choice of diagnostic line ratios although additional  sources  of uncertainty are introduced. 

The presence of the strong  \ciii\  emission line implies that  electron density \ne\ cannot be very high. However,  high density (\ne $ \sim 10^{11-13}$\cm3) is invoked to explain the  rich low ionization spectrum (especially \feii) observed in  quasars \citep[e.g., ][]{baldwinetal04}.  Several lines in the UV spectrum of I Zw 1 (prominent \feii, relatively strong \aliii, and  C{\sc iii}\l1176) point towards high density at least for the low ionization line (LIL) emitting zone  \citep{baldwinetal96,laoretal97a}. This low ionization BLR (LIL BLR) has very similar properties to the O{\sc i}  and Ca{\sc ii} emitting region identified by \citet{matsuokaetal08}. The region where these LILs  are produced cannot emit much \ciii\ if electron density exceeds 10$^{11}$ \cm3.  But is the \ciii\  line really so strong  in most quasars?  BLR conditions are certainly complex and the assumption of a single emitting region cannot explain both LILs and high ionization lines (HILs) \citep[][and references therein]{marzianietal10,wangetal11}.

In this paper we report  an analysis based on several diagnostic ratios used to constrain density, ionization parameter and metallicity in the BLR of two sources that are representative of the Narrow Line Seyfert 1 (NLSy1) sub-sample of quasars (\S \ref{sources}). We use methodological considerations that enable us to deblend and identify the principal lines of the spectra  (\S \ref{sec:considerations}). Our sources show weak  \ciii\ emission (relative to \siiii) simplifying interpretation of the emission line spectrum (\S \ref{sec:measurements}).  Diagnostic ratios are heuristically  defined (\S \ref{ratios})  and interpreted through an array of photo-ionization simulations (\S \ref{interp}). We show that they converge toward well-defined values of ionization and density (\S \ref{results}). We also consider the influence of  heating sources other than photoionization (\S \ref{alt}). Under the assumption of a pure photoionized gas, the present analysis can be used to determine the product of density and ionization parameter enabling us to estimate the distance of the BLR from the central continuum source (\rb) and the virial black hole mass (\mbh)\ (\S \ref{mass}). Finally, in \S \ref{conc} we give our conclusions.

\subsection{The Eigenvector 1 Parameter Space}
\label{sources}

 \citet{borosongreen92} (BG92) identified a series of correlations from principal component analysis (PCA) of the correlation matrix of emission line measures for a bright low-redshift quasar sample. The sample contained 87 PG quasars with $z \sim$ 0.5 (17 of them radio-loud, RL).  The first PCA eigenvector (hereafter E1) identified  correlations involving broad  \hb, broad FeII and narrow \oiiiopt\ emission lines.  In an effort to clarify the meaning of the E1, \citet{sulenticetal00a} searched for a correlation diagram that  showed maximal discrimination between the various Active Galaxy Nuclei (AGN) subclasses. The best E1 correlation space that we could identify involved measures of: 1) equivalent width of the \feii\l4570 blend (defined as the ratio \rfe = W(\feii\l4570)/W(\hb)) and 2) FWHM(\hb). These were supplemented with: 3) the soft X-ray photon index, \Gsoft\ and the centroid line shift of high ionization \civ. 
Figure 7 of \citet{sulenticetal00a} shows two-dimensional (2D) projections of this four-dimensional E1 (4DE1) space: FWHM(\hb) vs. \rfe,  \Gsoft\ vs. \rfe\ and FWHM(\hb) vs. \Gsoft. They supplemented the BG92 RL sample with an additional 18 sources (with comparable signal-to-noise S/N spectra) taken from Marziani et al. (1996) who reported W(\feii) measures over the  range 4240 -- 5850 \AA.  The  range 4240 -- 5850 \AA\ \feii\ flux  was divided by a factor $\approx$ 3.3 in order to obtain the flux in the range 4434-4684 \AA\ that was used  by BG92  (both works relied on  the same \feii\ template based on I Zw 1).  \citet{sulenticetal00a} separated the various subclasses of AGN into two populations: Population A-B separated at FWHM(\hb)= 4000 \kms. Physical drivers for the correlation were discussed in  \citet[e.g.][]{marzianietal03b} with black hole mass and Eddington ratio $L/L_{Edd}$ (where $L_{Edd} = 1.5 \times 10^{38} (M/M_\odot)$ is the Eddington Luminosity) identified as the principal drivers of change along the 4DE1 sequence. Black hole mass increases from Pop. A to Pop. B  while Eddington ratio decreases from Pop. A to Pop. B.

The division into two populations is, at the least, useful for highlighting major differences among Type 1 AGN, although spectral differences among objects within the same population are still noticeable, especially for Pop A sources (Figure 2 of \citealt{sulenticetal02}). This is the reason why they also divided the 4DE1 optical plane into bins of $\Delta$FWHM(\hb) = 4000 \kms\ and $\Delta$\rfe = 0.5. Bins A1, A2, A3, A4 are defined in terms of increasing \rfe,  while bins B1, B1$^+$, and B1$^{++}$ are defined in terms of increasing FWHM(\hb) (see Fig. 1 of \citealt{sulenticetal02}).  Sources belonging to the same spectral type show  similar spectroscopic measures and physical parameters (e.g. line profiles  and UV line ratios). Systematic changes are minimized within each spectral type so that an individual quasar can be taken as representative of all sources within a given spectral bin. The binning adopted in \citet{sulenticetal02} is valid for low $z$\ ($<$ 0.7) quasars. At higher $z$ an adjustment must be made since no sources with FWHM(\hb) $\textless$ 3500 \kms\ are found above redshift $z \sim$ 3 \citep{marzianietal09}.

\section{The targets}
\label{sources}

In this study we choose two representative examples of extreme Pop. A objects which show prominent \aliii\  and weak/absent \ciii\ emission lines. The objects are: the low redshift ($z = $ 0.06) NLSy1 prototype I Zw 1 
and the much more distant ($z = $ 3.23) SDSS J120144.36+011611.6 that appears to be a prototype of higher redshift NLSy1-like sources. Both are shown in Fig. \ref{fig:cont}. Sources like I Zw 1 are found at intermediate to high redshift.  Interpreted as the youngest and highest accreting sources (Sulentic et al. 2000) we might well expect to find more of them at high z.  SDSS J12014+0116 is a good example of a high-redshift, high-luminosity analog of I Zw 1 with broader lines (as can be seen in the right panels of Fig. \ref{fig:fits} where we show the line fits and can identify the broad component (BC) width of each object). In this paper, we shall use the acronym BC for this core or central component only. The spectrum of SDSS J12014+0116 shows lines that have FWHM(BC) $\sim$ 4000\kms\ which is much broader than the nominal NLSy1 cutoff of 2000\kms\ at low redshift. Emission line ratios (such as \aliii/\siiii, that can be derived from Table \ref{tab:obs}) and hence inferred physical conditions, are very close to those inferred for I Zw 1. This extends to other properties such as strong iron emission and  a large blue asymmetric/blueshifted component of \civ\ (extreme Pop. A objects in Sulentic et al. 2007). Thus the ``NLSy1 definition"  seems to be luminosity dependent (see also \citealt{netzertrakhtenbrot07}, \citealt{marzianietal09}) in the sense that we can extend this definition to high-luminosity and high redshift objects by extending the line widths limit to 4000 \kms\ and leaving all other properties unchanged \citep{dultzinetal11}.

A search in the SDSS DR7 for quasars in the redshift range where both \civ\ and the \l1900\AA\ blend are observed at optical wavelengths ($2 \ltsim z \ltsim 3$) yields more than 200 sources (out of 3000 candidates) with spectra resembling I Zw 1 on the basis of \aliii\  emission line strength (comparable to \siiii).  In practice, we performed  automatic measurements of the spectra, measuring a rough approximation of the \aliii/\siiii\ intensity ratio. As approximate as these measurements are, they are nonetheless suitable for identifying strong \aliii\ emitters.  SDSS J12014+0116, is a prototype of these strong \aliii\ emitters selected on the basis of moderate/high S/N. 
   
In summary, both at low- and high-$z$, as well as at low- and high-luminosity,  around 10\% of quasars are I Zw 1-like (i.e. NLSy1 type) on the basis of their emission line strengths.  Particularly strong \aliii\ emission is observed in SDSS J12014+0116 with almost the same intensity as \siiii. It also shows weak \ciii\ (discussed in \S \ref{sec:ciii}). In this paper we limit and justify  the application of our method to NLSy1s which comprise around 10\% of quasars. In a forthcoming paper, we shall address the applicability of our method to broader line quasars of both populations A and B.

\section{Observations and Data Analysis}
\label{anal}

\begin{figure}
\includegraphics[scale=0.9]{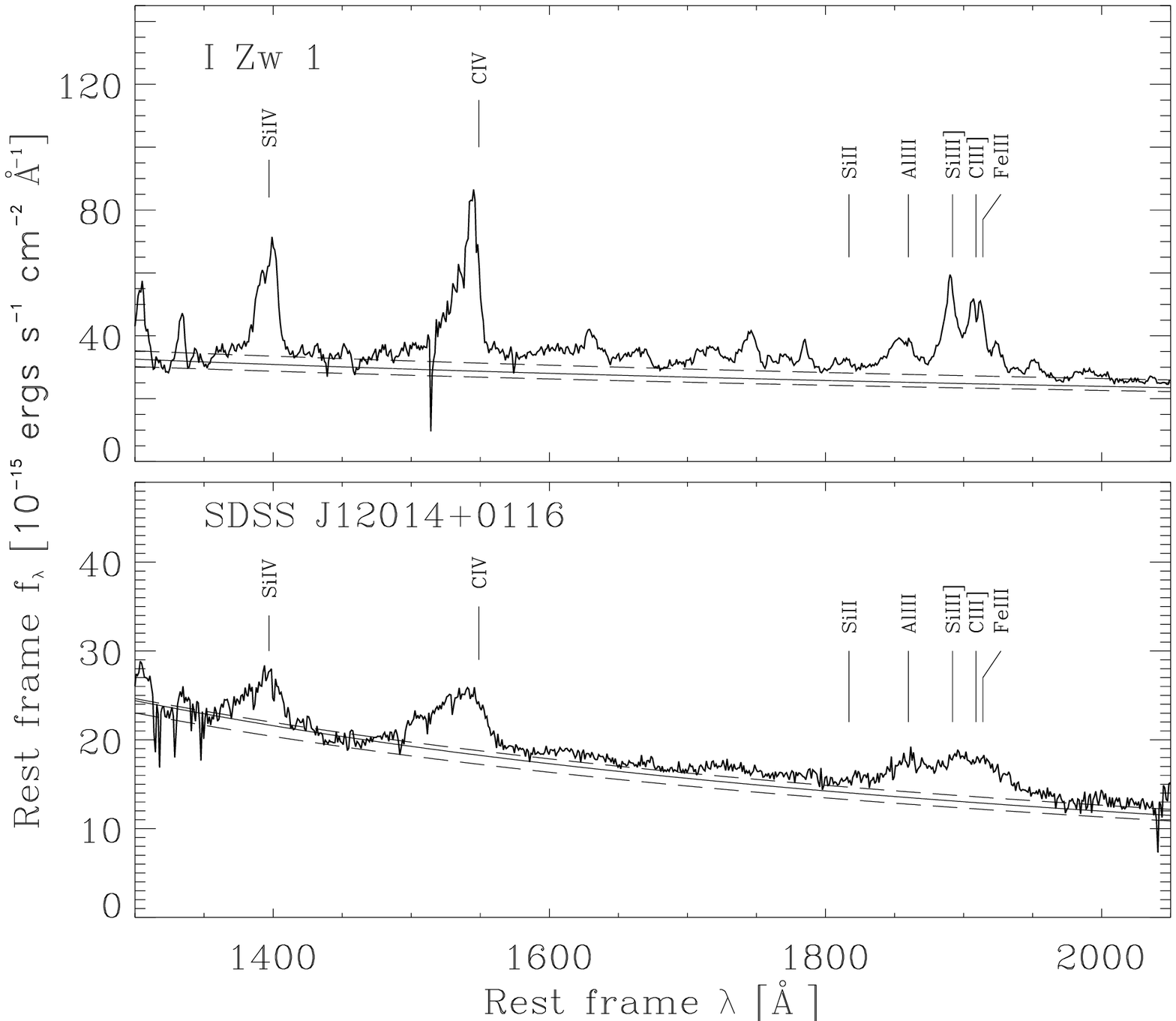}
\caption{Spectra of the two quasars in rest frame wavelength.  Upper panel:  I Zw 1; lower panel:  SDSS J1201+0116. Abscissa is rest frame in \AA, ordinate is specific flux in the rest frame in units 10$^{-13}$ ergs s$^{-1}$ cm$^{-2}$ \AA$^{-1}$. For both objects, the solid line shows the adopted continuum. Dashed lines under and above are the extreme cases. The principal emission lines are labeled.
\label{fig:cont} }
\end{figure}

We retrieve the UV spectrum of I Zw 1 (upper panel of Fig. \ref{fig:cont}), obtained with the FOS spectrograph, from the HST archives\footnote{It can be retrieved from the Web site at http://archive.stsci.edu}. This instrument had a spectral resolution of about 1300 over the 1150 \AA\ to 8500 \AA\ range.  The spectrum covers the range from 1150 to 3000 \AA\ with S/N of $\sim$45 around 1900\AA.  The SDSS J12014+0116 spectrum (lower panel of Fig. \ref{fig:cont}) covers the rest frame range from 1000 to 2100 \AA\  with S/N of $\sim$30 around 1900\AA. This spectrum was taken from the SDSS DR7 site within the the {\it Legacy} project\footnote{For more information about the {\it Legacy} survey, see the SDSS Web site at http://www.sdss.org}. The spectroscopy in this project covers a wavelength range from 3800  to 9200 \AA\ range with spectral resolution of 1800-2200.

\subsection{Methodological considerations}
\label{sec:considerations}

In order to deblend and identify the principal lines, as well as to extract the core of the broad emission lines,  we use the  following previous results:

\begin{itemize}
\item As mentioned above, \citep{sulenticetal02} divided Pop. A  and B sources into bins according to 
FWHM \hb\ and \rfe\ measures. In 2010, \citeauthor{zamfiretal10} computed median composite \hb\ spectra for 
each of the bins and showed that the broad \hb\ profiles in composite spectra of Population A sources are best fit by Lorentzian functions. In Population B objects, on the other hand, they are best described by Gaussian profiles (see also \citealt{marzianietal10}). This is an empirical result clearly shown in these works. Our sources are extreme Pop. A objects (bin A3) and thus we shall use Lorentzian profiles to model the broad components. 

\item \citet{marzianietal10} analyze 6 sources (including I Zw 1) representative of the six most populated bins of populations A and B. They  show that FWHM and profile shape of broad components of \siiii, \aliii\ and \civ\ are similar to those of \hb.  We do not have an \hb\ spectrum for SDSS J12014+0116 since there are no near IR data for this object. However for other  high z quasars, there are high S/N IR spectra  \citep{sulenticetal04} that show NLSy1-like sources (defined in \S \ref{sources}), with $M_B$ = -28 and FWHM(\hb) as much as 2000\kms\ broader than those with $M_B$ = -22. SDSS J12014+0116 shows $M_B$ = -29.8 with FWHM$\sim$4000\kms\ while for I Zw 1 $M_B$ = -23.5 and FWHM$\sim$2000 \kms. With larger samples at high redshift \citep[e.g.][]{marzianietal09} the result that the FWHM(\hb)  can be as high as 4000 \kms\ for NLSy1-like objects is confirmed. Following these results we use a Lorentzian profile with the same width (FWHM$\sim$4000 \kms\ in SDSS J12014+0116 and FWHM$\sim$2000 \kms\ in I Zw 1) for all the BCs of the broad emission lines.

\item HIL \civ\  profiles show significant differences between Pop. A and B sources.  In Pop. A. the peak of the line is often blueshifted and the profile blue asymmetric. We model this as a strongly blueshifted ($\la -1000$ \kms) \civ\ broad component (hereafter labeled BLUE; \citealt{sulenticetal07}) plus an unshifted broad component analogous to the one seen in \hb. We assume that the profile of this blueshifted component is Gaussian as discussed in  \citet{marzianietal10}.

 \item In Pop. A objects, low (\siii) and intermediate  (\aliii, \siiii) ionization lines  offer the simplification  of showing only the broad component associated with low-ionization emission \citep{marzianietal10}.
\end{itemize}

All the above results are taken into account for modeling the lines in this paper.

\subsection{Measurements}
\label{sec:measurements}

After identifying the emission lines needed for our study, we isolate the broad central component in each of them using spectral decompositions as explained below. We need the line fluxes to obtain \nh\ and $U$, the rest-frame specific flux at \l1700\AA\ to compute \rb\ and the FWHM of the broad components to estimate \mbh. We use the {\sc specfit}  {\sc IRAF} task \citep{kriss94} that enables us to fit the continuum,  emission and absorption line components, as well as \feii\  and \feiii\  contributions. We work with emission lines in the spectral range 1400 -- 2000 \AA\ which have been studied in great detail in both high and low-$z$ \ quasars and where identification of prominent resonance and inter-combination lines is well established.  

The steps we followed to accomplish identification, deblending and measurement of lines in each object were the following: 

\paragraph{1. The continuum.} We adopted a single power-law fit to describe it (Figure \ref{fig:cont}) using the continuum windows  around 1700 and 1280 \AA\  (see e.g. \citealt{francisetal91}).  \feii\ emission in these ranges is weak leading us to assume that continuum measurement in those windows is reliable enough for our method (within the uncertainties, see \S \ref{uncer}).

\paragraph{2. BC line widths and shifts.} We assume that a single value of FWHM (last Column of Tab.  \ref{tab:obs}) is adequate to fit the BCs of all lines. BC shifts of all lines are the same and consistent with the redshift reported in the Table. Due to the fact that the \ciii\ emission line is mostly emitted in a different region than the rest of the broad lines (see discussion in \S \ref{sec:ciii}) we do not impose the same restriction on the FWHM of \ciii.

 \paragraph{3. \l1900 \AA\ blend.} In Table \ref{tab:lineid} we summarize the properties of the strongest features expected that contribute to the \l1900\AA\ blend. Col. (1) lists the ion. Col. (2) is the rest frame wavelength. Col. (3) and (4) lists the the ionization potential and energy levels of the transition respectively. Col. (5) gives the configuration of the levels for the transition. Col. (6) and (7) give the transition probabilities and critical densities respectively. And in Col. (8) we give some notes for each ion. Forbidden lines of Si and C are not expected to be significantly emitted in the BLR and will not be further considered.

\feiii\ lines are frequent and strong in the vicinity of \ciii\ as seen in the SDSS template quasar spectrum   \citep{vandenberketal01}. They appear to be strong when  \aliii\ is also strong \citep{hartigbaldwin86}.  They are included in the photoionization simulations described below \citep{sigutetal04}.  \lya\ pumping enhances \feiii\l1914.0 (UV 34, \citealt{Johanssonetal00}) and this line can be a major contributor to the blend  on the red side of \ciii.  The spectrum of I Zw 1 convincingly demonstrates this effect: both \ciii\ and \feiii\l1914 are needed to account for the double peaked feature at 1910 \AA\ that is too broad to be explained by a single line (Fig. \ref{fig:fits}).\footnote{Note that the Galactic line of Mg I at 2026 \AA\ can contribute to the split appearance of the redshifted \l1910 \AA\ feature (Fig. \ref{fig:fits}). This absorption line is included in the {\sc specfit} analysis, but its resulting equivalent width is very small. The line separation between \ciii\ and \feiii\l 1914 is mostly intrinsic.}  We adopt the  template  (option B) of \citet{vestergaardwilkes01} plus additional \feiii\l1914, with the same profile as the other BCs lines, for modeling \feiii\ emission in our sources.  

\feii\ emission is not  strong in the spectral range we studied and the UV \feii\  template we adopt  is based on a suitable {\sc cloudy} simulation. Results on the \l1900\AA\ blend are not significantly affected by the assumed \feii\ contribution since it appears as a weak pseudo-continuum underlying the blend.  We explore maximum and minimum contributions of \feii\ by placing the highest and lowest possible  continua, as shown in Fig \ref{fig:cont}, and as explained below in \S\ref{uncer}. In Figure \ref{fig:fits} we show the contribution of \feii. If we increase or decrease this contribution, we will see an intensity variation of the strength of the lines, being \siii\ the most affected due its weakness  (see error bands in Fig. \ref{fig:contour}).  We take into account this strength variations for the error estimations.  

Then, for both objects, we sequentially model \feii\ and \feiii\ as preliminary steps. We anchor the \feii\ template to the 1785 \AA\ feature in order to normalize it. We continue with the fit of the \siii\ and \aliii\ emission lines that are fairly unblended. The main challenge is therefore to deconvolve \siiii, \ciii, and \feiii\l1914, noting that we will use only the less blended line, \siiii, for the eventual computation of diagnostic ratios. 
 
The next step is different for each object. The deblending of \siiii, \ciii, and \feiii\l1914 can be easily accomplished in the case of I Zw 1 because the lines are narrow. In the case of SDSS J12014+0116 the peak at \l1910 is consistent with \feiii\l1914 indicating that \feiii\ emission is dominating over \ciii. Thus, we first fit  \siiii\ and \feiii\l1914 to the observed peaks and the remaining part of the blend, as \ciii. We emphasize that in the \l1900\AA\ blend, the only two lines that are severely blended are \ciii\ and \feiii\l1914. This is more evident in the case of I Zw 1 because \ciii\ and \feiii\ have similar intensities. In the case of SDSS J12014+0116, \ciii\ is much weaker than \feiii\l1914 but still blended. We are not interested in the intensity of these two lines but only in a confirmation that \ciii\ is weak with respect to \siiii. In Figure \ref{fig:fitsdssciii} we show that this is valid even for the highest possible contribution of \ciii. The residuals in Figures \ref{fig:fits} and \ref{fig:fitsdssciii} reflect the noise. If we consider 1$\sigma$ above and below zero, then we say that a line is weak when it is below 1$\sigma$. For example, in the lower right panel of Fig. \ref{fig:fits}, \ciii\ is below 1$\sigma$, while \siii\ is around 1.5$\sigma$.

 \paragraph{ 4. \l1550 \AA\ feature.}  
As in the case of the previous blend, we need to keep in mind the complexity of this feature.
 In order to fit the HIL \civ, we have to take into account that it is decomposed into a Lorentzian BC with the same width and shifts of the intermediate ionization lines plus a blueshifted residual (assumed to be Gaussian in the {\sc specfit} procedure, discussed in \S \ref{sec:considerations}). So, in both objects, we fit first the \feii\ template with the same intensity as in the \l1900\AA\ blend. Then we fit the BC,  and looking at the residuals, we fit the BLUE component. Finally, we fit the underlying weaker emission lines \niv, \siiiuv\ and \heiiuv, when visible. We assume that the latter line has two components (BC and BLUE) with the same shift and width as \civ.   An equivalent approach has been successfully followed by several authors \citep{baldwinetal96,leighlymoore04,marzianietal10,wangetal11}.
 
We need to point out that in the case of I Zw 1, we observed a narrow component (NC) with a width of $\sim$800 \kms, close to the width to the BCs of the broad lines. The NC of \civ\ is observed in several low-z quasars (also radio loud) and Seyfert-1 nuclei, as well as in type-2 quasars \citep{sulenticetal07}.  Even if the line is collisionally excited (hence with an intensity proportional to the square of electron density), the larger volume of the narrow line region (proportional to ($R_{NLR}/R_{BLR})^{3}$) and the absence of collisional quenching at a relatively low density (as in the case of the \oiiiopt\ lines) make it possible to expect a significant \civ\ NC emission. We also observed that for I Zw 1, the NC of \civ\ is blueshifted. This is also observed in other narrow lines \citep{marzianietal10}, in agreement with expectations for the NLR of the extreme NLSy1s. For example, \oiiiopt\ is blueshifted with respect to \hb\ and to the systemic radial velocity of the host galaxy. In this way the analysis of the \civ\ NC is fully consistent with the  \oiiiopt\  and \hb\ analysis.

\paragraph{  5. \l1400 \AA\ blend.} This blend has been one of the most enigmatic features in quasar spectra \citep[e.g., ][]{willsnetzer79}. It is  known that the \siiv\ doublet is blended with \ion{O}{4} intercombination lines \citep{nussbaumerstorey82}. In our sources the \l1400\AA\ blend is very prominent, approximately 3-4$\times$ stronger relative  to \civ\ than in the SDSS composite quasar spectrum \citep{vandenberketal01}. This is consistent with the extreme metal enrichment we found in these sources  (\S \ref{chem}).
 
We are able to obtain  a reliable measurement of the \siiv\ doublet, even when we cannot measure all the  components of the \l1400\AA\ blend. Any \oiv\ contribution to the BCs is expected to be negligible \footnote{This is  not true for the BLUE component.  The physical properties derived for the BLUE component ($ 0 \gtsim \log U \gtsim -1$, $\log n \ltsim 9$, \citealt{leighly04,marzianietal10}) indicate that both \siiv\ and \oiv\ should contribute to the BLUE intensity. }.  We follow the same procedure to fit this blend as in \l1550\AA\ blend. Note that the \siiv\ low-ionization emission  line shows a double horned profile in I Zw 1  because of the large doublet separation and of the relatively narrow lines of this source. Since the peak at $\approx$ 1401 \AA\ is somewhat broader than the peak of the  individual component of \siiv, some \oiv\  emission might be associated to the \ion{O}{3}]\l1663\ emitting region. This is an additional, minority component that is needed to obtain a very good fit of the \l1400\AA\ blend. 

 In summary, the three following features were independently fitted (using {\sc specfit}, see Fig. \ref{fig:fits}):

\paragraph{ The \l1400 \AA\ blend} (whose profile is very similar to the one of \civ), with  BC of \siiv\ (we fit the doublet with individual lines at 1402 and 1394 \AA), and  one blueshifted component accounting for the contribution of both  \siiv\ and \oiv\ (+ semi-broad component most probably associated to \oiv\ in  I Zw 1). 
\paragraph{  The \l1550 \AA\ feature,} with \civ\ BC + BLUE + \feii\ + \ion{Si}{2}\l1531. We expect a contribution of \heii\ with a profile similar to \civ. In SDSS J12014+0116 we can see \heii\ BLUE. 
\paragraph{ The \l1900 \AA\ blend} considering \feii, \feiii, \siii, \aliii\ (we fit the doublet with independent lines, at 1855 and 1862 \AA, in Fig. \ref{fig:fits} we show only the sum), \siiii, \ciii\ and \feiii\l1914. This latter line is assumed to be an independent additional line of unknown intensity and with the same profile as the other BCs.

All BCs of the broad emission lines are assumed to have the same width and shift, leaving only their intensity as free parameters. The HIL blends involve mainly only two components.  As a result, the free parameters are reduced to the intensity of the BC lines (\ciii, \siiii, \aliii, \siii, \civ\ and \siiv), the intensity of the \feii\ and \feiii\ templates and the width and flux of the Gaussian minor components under \civ\ (\niv, \siiiuv, \heiiuv) and \siiv\ (\oiv). Table \ref{tab:obs} reports the fluxes of the BCs for intermediate  and high ionization lines. Col. (1) is the object name. Col. (2) is the redshift, for I Zw 1 we adopted the one reported in \citet{marzianietal10}; for the SDSS J12014+0116 object we use one reported in the SDSS database. Col. (3) is the rest-frame specific flux at \l1700\AA. Cols. (4) to (9) are the rest-frame line flux for the broad components only, and Col. (10) is the FWHM of all the broad components.

\subsubsection{Uncertainties}
\label{uncer}
The main sources of uncertainties in the measurements are the following:

\paragraph{\feii\  intensity (continuum placement).} Broad \feii\  emission can produce a pseudo-continuum affecting our estimates of emission line intensities.  \siii\ is especially affected in our spectra because it is weak (Fig. \ref{fig:cont}).  The effect is less noticeable for \civ\ and \siiv, since the expected \feii\  emission  underlying those lines is weak. The placement of this pseudo continuum also affects the determination of \rb\ for which we use the continuum flux measured at \l1700\AA\  (see column 3 of Table \ref{tab:obs} and equation \ref{eq:rblr1}).

\paragraph{ \feiii\ intensity.} These multiplets affect mainly the intensity of the \ciii\ emission line. We measured the maximum and minimum possible contributions of \ciii\ depending on the intensity of \feiii\l1914. The contribution of \ciii\ has no considerable effect to the intensity of \siiii\ even  in the case were \feiii\l1914 is maximum. This is important for the case of the SDSS object. In order to reproduce the observed \l1900\AA\ blend, if we increase the intensity of the \ciii\ emission line, the intensity of \feiii\l1914 necessarily has to decrease and viceversa. As  a result \siiii\ is not really affected by these variations. As  a result the line intensity of \siiii\ is  affected only by about $\sim$10\%, which is within the uncertainties, as shown in Table \ref{tab:obs}. In the lower right panel of Fig. \ref{fig:fits} and in Fig. \ref{fig:fitsdssciii}  we show the maximum and minimum contributions of \ciii\ and \feiii\l1914.

\paragraph{BLUE component.} In the case of the \civ\ and \siiv\ emission lines, the main source of error is the contribution of the BLUE component on the blue side of the central component. To a less extent, we may have a BLUE component contribution of \heii\ on the red side of \civ. In the previous section we describe how we deal with these contributions.

\paragraph{FWHM.}  When we run the {\sc specfit} routine we set the same FWHM for all the BCs of the broad emission lines. However, the routine introduces slight fluctuations around this initial value in order to obtain the best fit. When we vary the placement of the continuum, the FWHM determination is also affected. This source of error is reflected in the computation of the \mbh.

\subsubsection{\ciii\ emission}
\label{sec:ciii}
One must work carefully with the \l1900\AA\ blend because of the close proximity of the \ciii, \feiii\l1914 and \siiii\ emission lines. Both of our targets are  extreme Pop. A sources and one characteristic of this extreme population is that \ciii\ is weak or even absent (Fig. \ref{fig:fits}).  Extreme Pop. A sources show the lowest \ciii/\siiii\ ratio among all quasars in the E1 sequence \citep{bachevetal04}. 
We can see this effect in the upper right panel of Figure \ref{fig:fits} where we show the \l1900\AA\ blend for I Zw 1. The resolution is good enough to separate the peaks of \feiii\l1914 and \ciii. After fitting the \feiii\ template (including the \feiii\l1914 line), we see that in order to fit the observed spectrum the intensity of the \ciii\ line turns out to be comparable to \feiii\l1914. In the right  lower panel, for the SDSS  J12014+0116 object, the spectrum is noisier and the peaks of the lines of \ciii\ and \feiii\l1914 are not clearly seen. However, the observed peak is at the position of \feiii\l1914,  and if we follow the same procedure to fit the \feiii\ template in order to deconvolve the blend, it turns out to that the contribution of the \ciii\ emission line is practically insignificant. 
To estimate an upper limit to the \ciii\ line, we remove \feiii\l1914 (Fig. \ref{fig:fitsdssciii}). This estimate is \ciii\ $\approx$ 0.5 \siiii\ for the SDSS  J12014+0116 object, but the fit is poor on  the red side of the blend leaving a large residual.
In order to minimize the residual we added the maximum possible contribution of \ciii\ emission line (Fig. \ref{fig:fitsdssciii})
We can safely conclude that \ciii/\siiii\ $<$ 0.5 in this object. \ciii\ emission line in I Zw 1  is about $\approx$ 0.6 \siiii.  The very dense region emitting the LILs should produce no \ciii\ line because it is collisionally quenched; so any  emission from this line should arise in a different region. 

Available reverberation mapping results suggest that \ciii\ line is mainly emitted farther out from the central continuum source than some LILs and HILs. The results of reverberation mapping analysis are limited to a handful of low luminosity objects and cannot be generalized in a straightforward way. However, in these low luminosity objects  \ciii\  line  responds to continuum changes  on timescales much longer than \civ\ and other HILs. This results comes from the analysis of total \ciii\  + \siiii\ in NGC 3783 \citep{onkenpeterson02},  from the \ciii\ of NGC 4151 \citep{metzrothetal06} and of NGC 5548. It is intriguing that  the \ciii\ cross-correlation delay in  NGC 5548 (by far the best studied object) is even larger than that of \hb\ (32 ld vs. 20 ld; \citealt{petersonetal02,claveletal91}). For fixed density, lines of higher  ionization form at higher photon flux.  The C$^{++}$,   Al$^{++}$,  Si$^{++}$\  ionization potential are 24, 18, and 16 eV respectively. These  comparable ionization potentials  are well below the one of HILs, $X^{i -1} \ga$  50 eV. However, the much lower \ciii\  critical density  implies that the \ciii\ line should be  formed farther out than \siiii\ and \aliii\   if all these lines are produced  under similar ionization conditions. 

We also want to point out that there are  plausible  physical scenarios that can give rise to \ciii\ in a shielded, non-uniform environment where there gas has a range of densities at similar ionization parameter (i.e., the less dense gas is somehow shielded; \citealt{maiolinoetal10}).   Since we are considering  profiles integrated over the whole unresolved emitting region, the relevant question is however: how much will the \ciii\ emitting regions contribute to the lines we are considering?  The strongest \ciii\ emitters along the E1 sequence (spectral type A1) typically show \ciii/\siiii\ $\approx$\ 2.5. \citet{kuraszkiewiczetal04} more typically found  \ciii/\siiii\ $\approx$\  5.  The results of the fits indicate that \ciii\ emission is very weak respect to \siiii\ in our spectra. We obtain   \ciii / \siiii\ $\approx$\  0.6 for I Zw 1 and  \ciii / \siiii\ $\approx$\  0.2 for SDSS J12014+0116. Therefore the results of our procedure (described in \S \ref{sec:measurements})  will not be significantly affected. If the ionization parameter is $\log U \ltsim -2.5$, there will be modest HIL emission. Other intermediate ionization lines will be little affected. The results of the fits indicate that \ciii\ emission is low. We therefore neglect the effect of any \ciii\ emitting gas on the intermediate and high ionization lines. 

\section{Definition of Diagnostic Ratios and their Interpretation}

\subsection{Definition}

\label{ratios}

Our method for estimating \nh\ and $U$  involves several line ratios. We discuss the product \nh$\cdot U$ in the following sections and in \S\ref{mass} we use it to compute \rb. We define three groups of diagnostic ratios which should define density, ionization parameter, and metallicity for a given continuum shape and geometry. 

Line ratios such as  \aliii/\siiii\  are useful diagnostics over a range of density that depends on their transition probabilities. Emission lines originating from forbidden or semi-forbidden transitions become collisionally quenched above the critical density and therefore relatively weaker than lines for which collisional effects are still negligible.  The \aliii/\siiii\ ratio is well suited to sample the density range $10^{11} - 10^{13} $ cm$^{-3}$.   This corresponds to the densest emitting regions likely associated with production of LILs like the CaII IR triplet \citep{matsuokaetal08} and  Fe{\sc ii} \citep{baldwinetal04}.  

The  ratios \siii/\siiii\ and \siiv/\siiii\  are  sensitive to ionization, but roughly independent of metallicity since the lines come from different ionic species of the same element. Metallicity influences thermal and ionization conditions that in turn affect these ratios. The effect is however  second-order:  {\sc cloudy} simulation sets for different $Z$\ indicate that, for a fivefold increase in metal content the ratio \siiv/\siiii\ changes from $\approx$0.5 to $0.6$.

The ratio \siiv/\civ\ is mainly sensitive to the relative abundances of C and Si. The reason is that the ground and excited energy levels of these ions are very similar (the ionization potentials are close: 33.5 and 48 eV  for creation of Si$^{+3}$ and C$^{+3}$, respectively). This implies that the dependence on continuum shape and electron temperature of the ratio of this two resonance lines is small (see also \citealt{simonhamann10}).

\subsection{Interpretation}
\label{interp}

In order to illustrate how we employ {\sc cloudy}  simulations to estimate where the bulk of the line emission arises we refer to Figure \ref{fig:ionic_emissivity}. The left panel shows the ionic fraction as a function of geometrical depth in a slab (within a single BLR cloud) for which we choose a fixed column density\footnote{The value of \nc\  $= 10^{25}$ \cmq\ was needed to show that the Str\"omgren depth is less than the size of the clouds for the ionic states we are considering, i.e., that clouds are radiation bounded. 
In Figure \ref{fig:ionic_emissivity} we show that if we assume \nc\  $= 10^{25}$ \cmq,  the clouds become optically thick to electron scattering and we use them just to show the ionization structure and the line emissivity. Fig. 4 reproduces also the structure and emissivity of a typical log \nc $= 10^{23}$ slab (in this case the limit of  geometrical depth is log h $\approx$ 10.5).}
(\nc\  $= 10^{25}$ \cmq) and density (\nh $= 10^{12.5}$\cmq) exposed to a ``standard'' quasar continuum (the parameterization of \citealt{mathewsferland87}). It is customary to consider this parameterization as standard for the ionizing continuum. However,  that parameterization has been sometimes criticized and it is not unique, so that we also consider as an alternative the ionizing continuum parameterization of \citet{laoretal97a}. In this work, we use the term  `typical continuum' to designate the average of the two, with the two continua providing two extrema in number of ionizing photons expected at a specific flux measured on the non-ionizing part of the continuum. The `typical continuum' is the one used for our computation of \rb\ (see \S\ref{mass}).

Computing simulations at a fixed \nh\ and $U$\ values allows us to study how these parameters influence the diagnostic ratios we have defined. We make no assumptions about the geometry or kinematics of the BLR.  The slab of gas (i.e. a single cloud) might as well involve magnetically confined clouds \citep{reesetal89} or individual elements in an accretion disk atmosphere -- provided that photoionization is the only heating mechanism. 

Al$^{++}$, Fe$^{++}$, and Si$^{++}$\  are intermediate ionization lines sharing a region of dominance deep within a cloud -- where the HILs are also arise. It is therefore appropriate to consider intermediate-ionization line ratios between geometric depths $ h \sim 10^6 - 10^8$ cm (right panel of Figure \ref{fig:ionic_emissivity}). 
We prefer to show the product of the local line emissivity times the depth within the slab since the total line intensity is proportional to $\int \epsilon(h) dh$\ (in the absence of radiation transfer effects i.e., in a cloud where lines are optically thin),  and hence the product $\epsilon(h) \cdot h$\  gives a better idea of the total line emission at a distance $h$ than simply showing $\epsilon(h)$ vs. $h$.
 Fig. \ref{fig:ionic_emissivity} shows that  \ciii\ emission is expected to be orders of magnitude lower than the other lines and therefore likely undetectable.

Once  we have ascertained that the computed line ratios are physically consistent (i.e., all refer to a single emitting region except the ones involving \siii, whose intensity is significantly affected by a partially ionized zone (PIZ) emission), a multidimensional grid of {\sc cloudy}  simulations is needed to derive estimates for $U$, \nh\ and metallicity from  the spectral measurements \citep{ferlandetal98,koristaetal97}. {\sc cloudy} computes population levels of the relevant ionic stages of Si, C, Al and, especially for lines emitted in the fully ionized part of a gas slab (Fig. \ref{fig:ionic_emissivity}).  
Also, {\sc cloudy} is expected to be especially good for predicting intermediate and high ionization line fluxes that are produced in the fully-ionized region within the gas slab. Low ionization lines are produced in the PIZ by photoionization of hydrogen --primarily by soft X-ray photons. Since these X-ray photons create supra-thermal electrons and have a relatively low cross section for photoionization, the heating processes in the PIZ are inherently non-local, making the mean escape probability formalism used by {\sc cloudy} to treat radiation transfer a very rough approximation.

New simulations were needed since {\sc cloudy} has undergone steady and significant improvements since the time of \citet{koristaetal97}. The most relevant improvement is the addition of more ionic species in the simulations, with a 371-level of the Fe$^+$\ ion. This will make computation of equilibrium conditions more realistic even if, in the end, we expect that predictions of line intensity ratios will not be dramatically affected: iron is singly ionized only in the PIZ (Fig. \ref{fig:ionic_emissivity}, left panel).
Other relevant improvement for our study is the post-\citet{koristaetal97} update of the transition probability for \siii\ following \citet{callegaritrigueiros98} that changes the intensity of the line by a factor of $\sim$2 in the high density regime of the BLR.
We present the results of our simulations in Figure \ref{fig:contours}. Our results are not inconsistent with those obtained by  \citealt{koristaetal97}, for the same ranges of density and ionizing parameter: the overall behavior in the plane ($U$, \nh) of their Figure 3 is qualitatively consistent with the one derived from our simulations.

Simulations assume: 
(1) pure photoionization (see also \S \ref{alt}) and 
(2) spherical symmetry in continuum emission, i.e. that the ionizing continuum incident on a slab of gas of fixed density is similar to the observed continuum (see \S \ref{cav} for caveats concerning this assumption).

The ionization state of the gas will be mainly defined by the ionization parameter: 
\begin{equation}
\label{eq:u}
U = \frac {\int_{\nu_0}^{+\infty}  \frac{L_\nu} {h\nu} d\nu} {4\pi n_\mathrm{H} c r^2}
\end{equation}
where $L_{\nu}$\ is the specific luminosity per unit frequency, $h$\ is the Planck constant, $\nu_{0}$\ the Rydberg frequency, $n_H$ the hydrogen density,  $c$ the speed of light, and $r$ is the distance between the central source of ionizing radiation and the line emitting region.  

Simulations span the density range $7.00 \leq \log$ \nh$ \leq 14.00$, and $-4.50 \leq \log U \leq 00.00$, in intervals of  0.25 assuming  of plane-parallel geometry. We assume the standard value of \nc\ $= 10^{23}$ \cmq\ \citep[][and references therein]{netzermarziani10}.  
In Figure \ref{fig:contours} we show the isocontours for the ratios 
\aliii/\siiii, \siii/\siiii, \siiv/\siiii, \civ/\aliii, \civ/\siiii\ and \siiv/\civ\
derived from {\sc cloudy} simulations, for solar metallicity and ``standard'' quasar continuum, as parameterized by \citet{mathewsferland87} (see above). We considered three chemical compositions: (1) solar metallicity: (2) constant abundance ratio Al:Si:C with $Z=5 Z_{\odot}$; (3) an overabundance of Si and Al with respect to carbon by a factor 3, again with $Z= 5 Z_{\odot}$\ (5Z$_{\odot}$SiAl). This last condition comes from the yields listed  for type II Supernov\ae\ \citep{woosleyweaver95}. 
The Si overabundance is also supported by the chemical composition of the gas returned to the interstellar medium by an evolved population with a top-loaded initial mass function simulated using {\sc starburst99} \citep{leithereretal99}. The abundance of Al is assumed to scale with the one of Si (see also \S \ref{alt} on this assumption).

\section{Results}
\label{results}

\subsection{Density and Ionization Parameter}
\label{nhu}
In this section, we compute the  ratios analyzed above from the flux values and we estimate the physical parameters \nh\ and $U$ of the BLR, for I Zw 1 and  SDSS J12014+0116. We derived these values from the emission lines  measurements, reported in Table \ref{tab:obs}, and the {\sc cloudy} simulations of Figure \ref{fig:contours}.
Asymmetric errors due to the uncertainties described in \S \ref{uncer}, have been quadratically propagated following \citet{barlow03,barlow04}.  

In left panels of Figure \ref{fig:contour} we show the same \nh\ vs $U$ plane as in Figure  \ref{fig:contours} (in right panels, we use the 5Z$_\odot$SiAl simulation defined above). We choose only one isocontour for each ratio, and it is the one that corresponds to the measured value. There a is convergence of the contours lines defined by several crossing-points of the contour of the \aliii/\siiii\ ratio versus the contour of the other 5 ratios. This convergence defines the  \nh\ and $U$ values that point towards a low ionization plus high density range. We use the average value of all of the crossing-points to define a single \nh$\cdot U$ value, which is used to compute the \rb\ and the \mbh\ (\S \ref{mass}).

The uncertainties associated to a line ratio, $R$, broaden the lines defined by constant $R$ to a band limited by the ratios $R \pm \delta R$. In Fig. \ref{fig:contour}, we show them with bands. The largest of all uncertainties is related to \siii. This is the weakest line that we use and thus it is the line most strongly affected by the continuum placement. The value of \siii\ intensity is probably underestimated, and if it  was higher, the crossing point of \aliii/\siiii\ vs \siii/\siiii\ would be closer of the other crossing points, marked by an arrow in Fig. \ref{fig:contour}. We stress also that  \siii\ is the only low-ionization line considered in this study, and its formation is sensitive to the assumed X-ray continuum and other LIL-formation issues \citep{dumontmathez81,baldwinetal96}. 

The discrepancy  in the intersection point of diagnostic ratios in the plane (\nh, $U$) is significant for the \siiv/\civ\ ratio that depends mainly on the Si abundance relative to C. The  \l1400/\civ\ intensity ratio has been used as a metallicity indicator also by other authors \citep{juarezetal09,simonhamann10}. 
The discrepancy is less but still significant for the case when we consider the 5$Z_\odot$SiAl case. For this reason, we do not consider this ratio in the computation of the product of density and ionization parameter.

In similar  plots made for  $Z = 5 Z_{\odot}$SiAl (right panels of Fig. \ref{fig:contour}), the agreement of all the crossing-points improves:  the isocontour lines converge toward a better defined crossing-point. 

The high metallicity case $5Z_\odot$SiAl indicates higher $U$ and smaller \nh\  with respect to the case of solar abundances, if emission line ratios involving \civ\ are considered.  This reflects the increase in abundance of Si and Al relative to C with respect to solar: Si and Al lines appear stronger with respect to the \civ\ line because the elements Si and Al are more abundant, and not because a lower ionization level enhances  \siii, \siiii, \aliii\ emission with respect to \civ.

In the case of SDSS J1201+0116 we can see from Table \ref{tab:der} that the density is even higher than for I Zw 1, suggesting that \siiii\ is collisionally quenched  to make possible a rather high \aliii/\siiii\ ratio, $\approx 1$. The line intensity of \siiii\ is thus the result of the physical conditions of the emitting region, and in fact we measure a relatively low line intensity. The ratio \aliii/\siiii\ has a value of 0.6 for I Zw 1.

\subsection{The product ($U$\nh)}
\label{neu}

The products (\nh$\cdot U$) derived from the Figure \ref{fig:contour} and reported in Table \ref{tab:der} are marginally different for the two sources. It is intriguing, however, that while the  values of  \nh\ and $U$ taken separately depend significantly on metallicity, their product shows a weaker dependence: for $Z = 1 Z_{\odot}$,  we obtain $\log$(\nh$\cdot U$) $\approx$ 9.4 for I Zw 1 and 9.8  for SDSS J1201+0116. Similar results seem to hold also for a larger sample, including the objects monitored for reverberation mapping \citep{negrete11}. 

We also note that the $\log$(\nh$\cdot U$) values are very close to the results of several independent, previous studies  summarized in the Table \ref{tab:prevres}. 

\citet{baldwinetal96} presented a similar analysis. Their Fig. 2 organizes spectra in a sequence that is roughly corresponding to E1, going from \aliii- strong sources to objects whose spectra show prominent \ciii\ along with weak \aliii\ \citep{bachevetal04}.  The line components they isolated correspond to the ones we consider in this paper: a blue-shifted feature, and a more symmetric, unshifted and relatively narrow component that we call LIL-BC. They derive $\log$\nh$\approx$12.7\ and $\log U \approx -2.5$.  A somewhat lower density and higher ionization is indicated to optimize \feii\ emission, while the \caii\ IR triplet requires conditions that are very similar to the ones derived from the intermediate ionization lines.

We are not claiming that there is a single region that is able to account for all broad lines in all AGN. However,   a low ionization, high density region can account for \feii, \caii\ and the intermediate ionization line emission. This region is dominating the BC  of   emission lines (save weak \ciii) in the  objects considered in this paper, but apparently  becomes less relevant along the E1 sequence \citep{marzianietal10}.  The properties of   the emitting broad line region seem to be remarkably stable, keeping an almost constant $\log$(\nh$\cdot U$) (also seen in previous work, e.g. \citealt{wandeletal99}, \citealt{baldwinetal95}).  The issue is therefore whether we can apply the physical conditions (\nh\ and $U$) we deduce  to derive information of \rb\ and \mbh.

\section{Discussion }
\label{disc}

\subsection{Alternative Interpretations}
\label{alt}

 In this section we discuss other possible scenarios considering the influence of other sources of heating, besides  photoionization.
 We again remark that, on the basis of the analysis of section \S \ref{anal}, the use of intermediate and high ionization lines takes advantage of the most robust results of photoionization computations for AGNs. The complex issue of LIL formation is mostly avoided since we do not rely on the partially-ionized zone, whose physical properties may not be adequately modeled. Nonetheless,  results on density stem mainly  from the \aliii\ line being over-strong with respect to \siiii.  The observed \aliii/\siiii\  line ratio is inconsistent with density $\sim 10^{11}$ \cm3\ that would make  some \ciii\ emission possible.  Low ionization is inferred from the intrinsic weakness of \civ\ in these Pop. A sources with respect to \civ, that is observed in Pop. B objects \citep{sulenticetal07}.

While there is little doubt about the identification of the \aliii\ line doublet (a resonance line with large transition probability; in several NLSy1s the doublet is resolved with ratio $ \sim 1 - 1.2$, suggesting large optical depth), this line could be enhanced with respect to \siiii\ by some special mechanism. We can conceive two ways of  increasing \aliii\ with respect to \siiii: 
(1) a selective enhancement of the Al abundance with respect to Si, or 
(2) collisional ionization due to a heating mechanism different from photoionization. 

\subsubsection{Anomalous Chemical Composition}
\label{chem}

Evidence based on the  strength of the \nv\ line relative to the \civ\ and \heiiuv\ lines, indicates that   chemical abundances may be 5 to 10 times solar \citep{dhandaetal07} in high redshift quasars, with Z $\approx$ 5$Z_{\odot}$\ reputed  typical of high $z$\ quasars \citep{ferlandetal96}.  The [Si/C] enhancement (over solar) is supported by both  {\sc starburst 99} simulations \citep{leithereretal99} and supernova yields \citep{woosleyweaver95}.  
 
The production factors for  progenitors reported by \citet{woosleyweaver95} indicate that there should be little [Al/C] enhancement for 11 -- 20 \msol\ supernovae progenitors of solar metallicity of $Z_{\odot}$.  Massive progenitors are needed to raise the [Al/C] enrichment as they are the most efficient producers of aluminium. We integrate the production factors over a top-loaded mass function $\Phi(M) \propto M^{-x}$.
If $x = 1.3$\ (for $M \le 8 $ \msol, a value held canonical for the initial mass function (IMF) of young star forming systems), we obtain production factors ratios of  $\approx$ 3.0 and $\approx$ 2.5 for aluminium over carbon and silicon over carbon, respectively.  We also try changing the high-mass end of the IMF. With $x = 2.1$, we obtain a factor of $2$ enhancement for both aluminium and silicon.
 With $x = 1.1$,  we  obtain a significantly larger enhancement in aluminium than in silicon over carbon,  with production factor ratios $\approx$ 3.5 and $\approx$ 2.7. This condition is however rather extreme and the change in enrichment is rather small and will not affect significantly our diagnostic ratios.  

The observed values are more consistent with the assumption of a starburst. A factor of $\approx$ 3 enhancement of Si and Al over C indicates that BLR is made of gas whose chemical composition might reflect the enrichment due to a ``young'' starburst ($\ltsim {\rm few} \,10^{{7}}$ yr): a {\sc starburst 99} simulation  indicates that ejecta from a stellar system formed in an instantaneous burst should be enriched in Si in between 2 and 4$\cdot10^{{7}}$ yr.

\subsubsection{Mechanical heating and \feii\ emission}

NLSy1s of spectral type A3 and A4 are the sources for which pure photoionization models are  deficient as far as the prominence of \feii\ emission is concerned \citep{jolyetal08}. Since  \aliii\ prominence correlates with \feii\ prominence along the E1 sequence, there might be a mechanical heating contribution to the thermal and ionization balance that is often invoked to explain \feii\ emission \citep{collinjoly00,baldwinetal04}.  This second possibility reopens the issue of LIL formation that is too complex to be discussed in the present paper.  

We can consider whether gas in collisional equilibrium at a fixed electron temperature could give rise to a spectrum accounting for the strong \feiiq\ and \aliii\ emission lines, as well as the line ratio \siiii/\ciii\ $\approx$ 2 as observed in I Zw 1.  We cannot ignore, however, that Balmer line emitting gas appears to be pre-eminently photoionized, also in NLSy1 objects, as convincingly demonstrated by several reverberation mapping campaigns, and that the width of \feii\ is consistent with the width of the \hb\ rms spectrum \citep{sulenticetal06a}. Continuity arguments along the E1 sequence and monitoring studies  indicate that photoionization cannot be fully dismissed. Also, I Zw 1 is among the strongest emitter along the E1 ``main sequence'' of \citet{sulenticetal00a}, but not an ultrastrong \feii\ emitter  as defined by \citet[][]{liparietal93}. Ultrastrong \feii\ emitters are outliers in the E1 optical plane \citep{sulenticetal06}, and often ultra-luminous IR galaxies (ULIRGs) and extreme broad absorption line (BAL) sources. It is legitimate to suspect very different physical conditions in that case.    

There is convincing evidence that \feii\ is responding to continuum changes, although measurements are very difficult and the response of \feii\ might be more erratic than \hb\ \citep{vestergaardpeterson05,peterson11}. PG 1700$+$518 is a strong \feii\ emitter, and monitoring of \feiiopt\ indicates a response on a timescale consistent with the one obtained for \hb\ \citep[][note that \citealt{bianetal10} are not able to derive a reverberation radius for \hb, unlike previous monitoring campaigns]{bianetal10,petersonetal04}. On the converse Akn 120, a source with significant \feii\ \citep{marzianietal92,korista92} shows a response that may be consistent with a region more distant than the one of \hb, or with a region  that is not photoionized \citep{kuehnetal08}.  A recent work suggests that the large range of observed \rfe\ values can be explained by photoionization, if the variation of iron abundance  in  dusty gas is taken into account \citep{shieldsetal10}. 

The (\nh, $U$) solution we find falls below the ionization level that maximizes \feiiq\ emission. Adopting the suggestion of S. Collin and collaborators \citep[e.g.][]{joly87}, a simulation computed in the case of a very weak photoionizing continuum and dominance of collisional ionization at T$= 7000$K, would  enhance the \rfe\ ratio to levels even in excess to the one observed, leaving however unaffected the intermediate and high ionization lines (since there are few electrons with energies sufficient to ionize their parent ionic species). 
 
A system dominated by collisions at a fixed, single temperature higher than $\approx 10000$K  would yield contradictory  results.  For example, a collisional equilibrium solution at T$= 20000$K and $\log$\nh = 10 would imply \rfe\  exceeding by a factor 2 the observed value. Ratios involving the \aliii, \siiii\ and \siiv\ emission lines would be consistent with the observed ones, but the solution overpredicts the intensity of intermediate ionization emission lines  by a  factor $\sim 100$, with little \ciii\ and no \civ. If any such region exists, it must contribute little to \caii\ and \feii\ emission.  Another paradox of this case would be that \siiv\ and \civ\ require different $T$ to account for their intensity ratio. Reverberation mapping of Seyfert nuclei indicates however that the two lines are most close in response times \citep{koristaetal95,wandersetal97}, which is thus inconsistent with a different temperature of the emission line gas.

If the only and dominant source of ionization was mechanical deposition of energy due to shocks and/or friction (shear), our analysis based on photoionization would not be valid. Clearly, an  {\em ad hoc} solution can be found  invoking a range of temperatures. However, considering the low EW of all lines in the  sources of this paper, and the continuity with the other sources in the E1 sequence, we conclude  that there is no convincing evidence that shocks are dominating the emission of HIL and intermediate ionization lines.  An additional heating source might significantly affect only the low-$T$ PIZ where most \feii\ is emitted \citep{collinjoly00}.

\subsection{Implications}

\label{mass}

In \S \ref{results} we estimate the  product \nh$\cdot U$ and now we can compute the distance of the BLR
(\rb) from the central continuum source and the black hole mass (\mbh). They are key parameters 
that allow us to better understand gas dynamics in the emitting region as well as quasar phenomenology and evolution. The dependence of $U$\ on \rb\ was used to derive black hole masses assuming a plausible  average  value of the product \nh $\cdot U$. We also use FWHM(\hbbc) under the assumption that the BC arises 
from a virialized medium \citep{padovanietal90,wandeletal99}.  Eq. \ref{eq:u} can be rewritten as
\begin{equation}
r_{BLR} = \left[ \frac {\int_{\nu_0}^{+\infty}  \frac{L_\nu} {h\nu} d\nu} {4\pi n_\mathrm{H} U c} \right]^{1/2}
\end{equation}
and also as
\begin{equation}
\label{eq:rdp}
r_{BLR} = \frac 1{h^{1/2} c} (n_\mathrm{H} U)^{-1/2} \left( \int_{0}^{\lambda_{Ly}} f_\lambda \lambda d\lambda \right) ^{1/2} d_\mathrm{C} \label{eq:rblr}
\end{equation}
where $d_C$ is the the total line-of-sight comoving distance \citep{hoggfruchter99}:
\begin{equation}
d_\mathrm{C} = \frac{c}{H_{0}} \zeta(z, \Omega_M, \Omega_\Lambda)= \frac{c}{H_{0}}  \int_{0}^{z} \frac{dz'}{\sqrt{\Omega_M(1+z)^3+\Omega_\Lambda}},
\end{equation}
where we adopt the Hubble constant  $H_{0}$ = 70 km s$^{-1}$ Mpc$^{-1}$. The function $\zeta$ has been interpolated as a function of redshift:
\begin{equation}
\label{zeta}
\zeta(z, \Omega_M, \Omega_\Lambda)  \approx \left[1.500\left(1-e^{-\frac{z}{6.107}}\right)+0.996\left(1-e^{-\frac{z}{1.266}}\right)\right].
\end{equation}
with $\Omega_M = 0.3$ and $\Omega_\Lambda = 0.7$, given by \citet{sulenticetal06}.  

In equation \ref{eq:rdp}, we transformed the integral from units of frequency to wavelength and rewrote the expression for \rb\ in terms of the {\em rest frame} specific flux $f_{\lambda}$\ that can be easily derived from the observed flux, namely:
\begin{equation} 
L_{\nu} \nu^{-1} d \nu = 4 \pi c^{-1} d_\mathrm{L}^{2} f_{\lambda}  \lambda d \lambda
 \end{equation}
with $d_L$ the luminosity distance 
that is related to $d_C$ by the formula $d_L=d_C(1+z)$. 
   
Note that
\begin{equation}
\int_{0}^{\lambda_{Ly}} f_\lambda \lambda d\lambda =  f_{\lambda0}  \cdot \tilde{Q}_\mathrm{H}, \quad \mathrm{with} \quad  \tilde{Q}_\mathrm{H} = \int_{0}^{\lambda_\mathrm{Ly}} \tilde{s}_\lambda \lambda d\lambda,
\label{eq:Q_H}
\end{equation}

where $\lambda_0 = 1700$\AA. $\tilde{Q}_{H}$ depends on the shape of the ionizing  continuum for a given specific flux with the integral carried out from the Lyman limit to the shortest wavelengths. 
We use $\tilde{s}_\lambda$ to define the Spectral Energy Distribution (SED) following \citet{mathewsferland87} and \citet{laoretal97b} conveniently parameterized as a set of broken power-laws. $\tilde{Q}_H$ is $\approx 0.00963$ cm\AA\  in the case  of the \citet{laoretal97b} continuum and $\approx 0.02181 $ cm\AA\  for \citet{mathewsferland87}. We use an average value because the two SEDs give a small difference in estimated number of ionizing photons (see below).

Expressing \rb\ in units of light-days, and scaling the variables to convenient units, Eq. \ref{eq:rblr} becomes:
\begin{equation}
r_{\rm BLR} \approx 93 \left[ \frac{f_{\lambda_0,-15} \tilde{Q}_\mathrm{H,0.01}}{(n_{\mathrm H} U)_{10}} \right]^\frac{1}{2} 
\zeta(z, 0.3, 0.7)  ~~\mathrm{light \, days.} 
\label{eq:rblr1}
\end{equation}
In this equation, $f_{\lambda_0,-15}$\ is the specific rest frame flux (measured on the spectra) in units of 10$^{-15}$ erg s$^{1}$ \cmq\ \AA$^{-1}$, $\tilde{Q}_{H,0.01}$ is normalized to $10^{-2}$ cm\AA. The product \nh$U$\ is normalized to 10$^{10}$ cm$^{-3}$ , $\zeta(z, 0.3, 0.7)$ is derived from equation \ref{zeta} and \rb\ is now expressed in units of light days.

Knowing \rb\ we can calculate the $M_\mathrm{BH}$ assuming virial motions of the gas
\begin{equation}
\label{eq:vir}
M_\mathrm{BH} = f \frac{\Delta v^2 r_\mathrm{BLR}}G.
\end{equation}
or,
\begin{equation}
M_\mathrm{BH} = \frac 3{4G} f_{0.75} (FWHM)^2 r_{BLR} 
\end{equation}
with the geometry term $f \approx 0.75$, corresponding to $f_{0.75} \approx 1.0$  \citep{grahametal11}.  The factor $f$ depends on the details of the geometry, kinematics, and orientation of the BLR and is expected to be of order unity. This factor converts the measured velocity widths into an intrinsic Keplerian velocity (\citealt{petersonwandel00}, \citealt{onkenetal04}, \citealt{grahametal11}).

Resultant \rb\ and \mbh\ estimates are reported in  Table \ref{tab:der}. Errors in this Table are at 2$\sigma$\ confidence level. They are not symmetrical around this value. They were propagated quadratically, following \citet{barlow03,barlow04}. We consider three sources of uncertainty  in the \rb\ computations:

1. The error in the determination of \nh\ and $U$ that is described in detail in \S \ref{nhu}. We use the average value for all of the crossing-points in Figure \ref{fig:contour} to define a single  \nh$\cdot U$ value which is used in equation \ref{eq:rdp} to compute the \rb.

2.  The error derived from the shape of the ionizing continuum, which is used in the computation of the \rb. The two SEDs that we assumed as extreme yield a difference in ionizing photons of a factor 0.17 dex. At a 2$\sigma$\ confidence level this corresponds to an uncertainty in the number of ionizing photons of $\pm$ 0.057 dex. 

3.  Errors in the specific fluxes (at \l1700\AA\ Col. 3 of Tab. \ref{tab:obs}), intrinsic to the spectrophotometry. We also consider the error in the continuum placement (discussed in \S \ref{uncer}).

In the determination of \mbh\ we consider two sources of error.

1. The combined error of the three sources of uncertainties on the \rb\ computation described above.

2. The error on the determination of the FWHM, that is discussed at the end of the section \ref{uncer}.

Using our derivations for \nh\ and $U$ we obtain  the \rb\ and \mbh\ values listed in Table \ref{tab:der}.  The last two columns give virial black hole masses following our method and using the \mbh\ -- luminosity correlation from \citet{vestergaardpeterson06} respectively. Notice that \citet{vestergaardpeterson06} used a different value for the geometry term which is $f$ = 1.4 (\citealt{onkenetal04} and \citealt{wooetal10}), and the implication in the mass computation is that our masses are a factor of two lower.  This was taken into account making the correction $log(f_{ours}/f_{V\&P}) = log(0.75/1.4) = -0.27$. We subtract this quantity to the \citet{vestergaardpeterson06}  \mbh\ formula used to compute the masses in column 7 of Table  \ref{tab:der}.

The intrinsic dispersion in the \citet{vestergaardpeterson06} relation is $\pm 0.66$  dex in black hole mass at a $2\sigma$\ confidence level. We are not considering here the scatter in the relation \rb\ -$L$ derived by \citet{bentzetal09} which involves sources where \rb\ was derived from reverberation mapping. The scatter in the \rb\ -$L$ correlation is $\approx$ 0.2  dex in \rb. These \rb\ determinations are, in spite of many caveats recently summarized in \citet{marzianisulentic12}, probably the best ones available. \citet{kaspietal07} were able to derive one point in the \rb\ -$L$ correlation directly from reverberation mapping of a high-$z$ (2.17) object. Additional observations will allow to check whether the Kaspi empirical relationship holds at high redshift. At the moment we cannot rely on the mass derivation of one source only.
The  UV extrapolation for single-epoch virial-broadening estimates in \citet{vestergaardpeterson06} shows a much larger scatter but is the only relation that provides \mbh\ a suitable comparison with our result. Errors obtained with our method are a factor of $\sim$3 smaller than the dispersion in the \citet{vestergaardpeterson06} relation. However, the latter errors are statistical making the comparison not very straightforward. In a forthcoming paper (Negrete et al. 2012 in preparation) we will give the results of a statistical analysis applying our method to a larger number of quasars than reported here (8 at high-$z$ and 14 at low-$z$, belonging to both populations A and B). Preliminary results are given in Negrete (2011). These are all Type 1 (broader line) quasars.

\subsubsection{Caveats}
\label{cav}

 Our estimation of  \rb\ assumes that the continuum incident on the line-emitting gas is ``as observed'' by us. \citet{leighly04} suggests (for two IZw1-like sources: IRAS 13224-3809 and 1H0707-495) that a high-ionization wind producing the BLUE component (we fit to the HILs: Marziani et al. 2010) intercepts most of the ionizing flux -- leaving only a small fraction available for photoionizing the LIL emitting region. The empirical analysis of 
their UV spectra is similar to ours with separation of an unshifted BC and BLUE components  \citep{leighlymoore04}.   While continuum absorption  should not  strongly influence the determination of \nh\ and $U$\ (they are set by line ratios dependent on physical conditions in the gas) the \rb\ value will be affected (Eq. \ref{eq:rblr1}). Absorption as extreme as hypothesized by \citet{leighly04} (a factor 10) would lead to  a decrease of \rb\ by $\approx$0.5\ dex. However, any consideration is highly dependent on the assumed geometry.  The configuration envisaged by \citet{leighly04}  suffers from an immediate observational difficulty. If the LIL emitting regions were being hit by a wind then the LILs should be emitted by gas ablated from a thick structure and with some of the wind momentum  transferred to the ablated gas. There is no observational (kinematical) evidence for this: the LILs and intermediate-ionization lines show stable, unshifted Ê(to within a few hundreds \kms) and symmetric profiles. It is worth noting that the LILs are even unaffected by the presence of a powerful radio jet in radio-loud quasars \citep{marzianietal03b}. In summary, current evidence suggests that the LIL emitting region is not strongly affected by quasar outflows.

\section{CONCLUSION}
\label{conc}

Diagnostic line ratios for estimation of density, ionization and metallicity can be found and exploited for  NLSy1-like sources at high redshift. Accurate diagnostics require high S/N and moderate dispersion spectra but in principle can be applied to very high $z$ ($>$ 6.5) using data from IR spectrometers.  The product  (\nh$\cdot U$)  yields the possibility of deriving \rb\ and \mbh\ for a significant number of sources for comparison with estimates derived from extrapolation of the Kaspi relation. Negrete et al. (2012, in preparation) will present an analysis of the applicability of the photoionization method described here to the general population of quasars  starting from the sources with reverberation mapping determinations of \rb. 

It is important to stress however, that the line deblending that allow us to obtain the line fluxes used to compute \nh\ and $U$, is not a trivial task. There is a wide diversity in the quasar spectra and one must consider previous results that allow us to predict and/or expect the presence and shape of various components in the broad lines  (as discussed in \S\ref{sec:considerations}). One also has to take into account expectations on certain line ratios (e.g. the extreme Pop. A sources show the lowest \ciii/\siiii\ ratio among all quasars in the E1 sequence, see \S \ref{sec:ciii}).
 
The results of this study are preliminary in many ways. Several issues remain open.
1) Is there a ``universal'' \nh$\cdot U$ product that can be used for all AGN? The consistency among our results, the ones reviewed in \S \ref{neu} and the \rb-$L$ relation, suggest that this might be the case, at least to a first approximation and especially for Population A sources.  2) A grid of simulations could be refined considering more intermediate metallicity cases. 3) Application of this method has been carried out without considerations about the geometry and kinematics of the BLR. A more refined treatment should consider likely scenarios and investigate their influence on derived physical parameters. 4) Considerations of geometry and kinematics could lead to a physical model accounting for non-thermal heating and production of \feii\ emission in a context more appropriate than that of pure photoionization.

As a final remark, we stress that values of \mbh\ derived from photoionization considerations, even assuming an average (\nh$\cdot U$) are probably more accurate than the ones derived from the mass-$L$\ relation. The main reason is that we are using each individual quasar luminosity (and not a correlation with large scatter) and the properties of a region that remains similar to itself over a wide range of luminosity.

\acknowledgments
D.Dultzin acknowledges support form grant IN111610 PAPIIT UNAM. 
Funding for the SDSS and SDSS-II has been provided by the Alfred P. Sloan Foundation, the Participating Institutions, the National Science Foundation, the U.S. Department of Energy, the National Aeronautics and Space Administration, the Japanese Monbukagakusho, the Max Planck Society, and the Higher Education Funding Council for England. The SDSS Web Site is http://www.sdss.org. The SDSS is managed by the Astrophysical Research Consortium for the Participating Institutions listed at the http://www.sdss.org.

\clearpage

\begin{deluxetable}{ccccccccccc}
\tabletypesize{\scriptsize}
\tablecaption{Line Components in the $\lambda$1900 blend\label{tab:lineid}}
\tablewidth{14cm}
\tablehead{
\colhead{Ion} & \colhead{$\lambda$} & \colhead{$X$} &\colhead{$E_l - E_u$} & 
\colhead{Transition} & \colhead{$A_{ki}$} & \colhead{$n_\mathrm{c}$} &  \multicolumn{4}{c} {Note} \\
\colhead{} & \colhead{\AA} & \colhead{eV} &\colhead{eV} & 
\colhead{} & \colhead{s$^{-1}$} & \colhead{cm$^{-3}$} &\colhead{}&\colhead{}&\colhead{}&\colhead{}\\
\colhead{(1)} & \colhead{(2)} & \colhead{(3)} & \colhead{(4)} & \colhead{(5)} & \colhead{(6)} & \colhead{(7)} &\colhead{}& \colhead{(8)} 
}
\startdata
Si II	&	1808.00	&	8.15		&	0.000	-	6.857	&	${}^2 D^o_{3/2} \rightarrow {}^2 P_{1/2}$	&	$2.54 \cdot 10^6$	&	\nodata	& 1  	\\
Si II	&	1816.92	&	8.15		&	0.036	-	6.859	&	${}^2 D^o_{5/2} \rightarrow {}^2 P_{3/2}$	&	$2.65 \cdot 10^6$	&	\nodata	& 1		\\
Al III	&	1854.716	&	18.83	&	0.000	-	6.685	&	${}^2 P^o_{3/2} \rightarrow {}^2 S_{1/2}$	&	$5.40 \cdot 10^8$	&	\nodata	& 1		\\
Al III	&	1862.790	&	18.83	&	0.000	-	6.656	&	${}^2 P^o_{1/2} \rightarrow {}^2 S_{1/2}$	&	$5.33 \cdot 10^8$	&	\nodata	& 1		\\

\[[Si III]&	1882.7	&	16.34	&	0.000	-	6.585	&	${}^3 P^o_2 \rightarrow {}^1 S_0$		&	0.012			&	$6.4 \cdot 10^{4}$	& 1&2&3	\\
Si III]	&	1892.03	&	16.34	&	0.000	-	6.553	&	${}^3 P^o_1 \rightarrow {}^1 S_0$		&	16700			&	$2.1 \cdot 10^{11}$	& 1&4&5	\\

\[[C III]&	1906.7	&	24.38	&	0.000	-	6.502	&	${}^3 P^o_2 \rightarrow {}^1 S_0$		&	0.0052			&	$7.7 \cdot 10^{4}$	&  1&2&6 	\\
C III]	&	1908.734	&	24.38	&	0.000	-	6.495	&	${}^3 P^o_1 \rightarrow {}^1 S_0$		&	114				&	$1.4 \cdot 10^{10}$	&  1&2&4&5	\\

Fe III	&	1914.066	&	16.18	&	3.727	-	10.200	&	$z^7 P^o_3 \rightarrow a^7 S_3$		&	$6.6 \cdot 10^8$	&	   \nodata	&	7	\\
\enddata
\tablewidth{13cm}
\tablecomments{All wavelengths are in vacuum. \\ (1)  Ralchenko, Yu., Kramida, A.E., Reader, J.,  and NIST ASD Team (2008). NIST Atomic Spectra Database (version 3.1.5). Available at: http://physics.nist.gov/asd3.  \\ (2) Feibelman \& Aller (1987). \\ (3) $n_{\mathrm{c}}$\ computed following Shaw \& Dufour (1995). \\ (4)  Morton (1991).  \\ (5) Feldman (1992).  \\ (6)  Zheng (1988).  \\ (7) Wavelength and $A_{ki}$\ from Ekberg (1993), energy levels from Edl\'en and Swings (1942). }
\end{deluxetable}

\bigskip

\begin{deluxetable}{lcccccccccc}
\tabletypesize{\scriptsize}
\setlength{\tabcolsep}{1pt}
\tablecaption{Measured quantities \label{tab:obs}}
\tablewidth{0cm}
\tablehead{
\colhead{Object} & \colhead{$z$} & \colhead{$f_{\lambda}$(1700 \AA)\tablenotemark{a}} & \colhead{\siiv\tablenotemark{b}} & \colhead{\civ\tablenotemark{b}} & \colhead{\siii\tablenotemark{b}} & \colhead{\aliii\tablenotemark{b}} & \colhead{\siiii\tablenotemark{b}} &
\colhead{\ciii\tablenotemark{b}} & \colhead{FWHM\tablenotemark{c}} \\
\colhead{(1)} & \colhead{(2)} & \colhead{(3)} & \colhead{(4)} & \colhead{(5)} & \colhead{(6)} & \colhead{(7)} & \colhead{(8)} & \colhead{(9)} & \colhead{(10)} }
\startdata
I Zw 1                          & 0.0605  & 2.7$^{+0.3}_{-0.2}$ & 27.4$^{+6.0}_{-3.5}$ & 28.4$^{+5.4}_{-4.6}$  & 5.4$^{+3.7}_{-3.0}$ &18.9$^{+2.9}_{-3.3}$ & 31.1$^{+5.5}_{-5.7}$ & 19.0$^{+2.5}_{-3.5}$ & 1050$\pm$200 \\
SDSS J1201+0116 & 3.2332  & 1.6$\pm$0.2  & 14.3$^{+2.6}_{-3.2}$  & 15.5$^{+4.3}_{-2.5}$ & 5.5$^{+3.0}_{-3.2}$ & 11.9$^{+2.4}_{-2.3}$  & 12.7$^{+2.4}_{-1.8}$ & 2.6 $^{+3.2}_{-1.8}$ &    4000$\pm$400
\enddata
\tablenotetext{a}{Rest-frame specific continuum flux   at 1700 \AA\ in units of 10$^{-14}$ erg \, s$^{-1}$ cm$^{-2}$ \AA$^{-1}$.}
\tablenotetext{b}{Rest-frame line flux of intermediate ionization line BC and of the \civ\ BC in units of 10$^{-14}$ erg \, s$^{-1}$ cm$^{-2}$.}
\tablenotetext{c}{Rest frame FWHM of intermediate ionization line BC and of the \civ\ BC in units of \kms.}
\end{deluxetable}
\bigskip

\begin{deluxetable}{llcccccccc}
\tabletypesize{\scriptsize}
\setlength{\tabcolsep}{1pt}
\tablecaption{Derived quantities \label{tab:der}}
\tablewidth{0cm}
\tablehead{
\colhead{Object} & \colhead{$\log$\nh\tablenotemark{a}} & \colhead{$\log U$} &\colhead{$\log$(\nh$U$)}    & \colhead{$\log$\rb\tablenotemark{b}} & \colhead{$\log$\mbh\tablenotemark{c}} & \colhead{$\log$\mbh(VP06)\tablenotemark{d}}\\
\colhead{(1)} & \colhead{(2)} & \colhead{(3)} & \colhead{(4)} & \colhead{(5)} & \colhead{(6)} & \colhead{(7)} }
\startdata
I Zw 1                         & 12.00$^{+0.32}_{-0.23}$ & -2.65$^{+0.21}_{-0.11}$ & 9.35$^{+0.33}_{-0.23}$ & 17.30$^{+0.17}_{-0.12}$ & 7.30$^{+0.23}_{-0.19}$ & 6.70 \\
SDSS J1201+0116 & 12.63$^{+0.28}_{-0.24}$ & -2.79$^{+0.12}_{-0.06}$ & 9.84$^{+0.28}_{-0.23}$& 18.31$^{+0.14}_{-0.12}$ & 9.39$^{+0.17}_{-0.15}$  & 9.29
 \enddata
\tablenotetext{a}{\nh\ in units of \cm3.}
\tablenotetext{b}{\rb\ in units of cm.}
\tablenotetext{c}{\mbh\ in units of \msol\ computed with the FWHM values of Table \ref{tab:obs}.}
\tablenotetext{d}{\mbh\  $\pm 0.66$  dex at a $2\sigma$\ confidence level, in units of \msol\  computed following \citet{vestergaardpeterson06}, and input parameters reported in note (b). See text for details.} 
\end{deluxetable}

\begin{deluxetable}{llcccccc}
\tabletypesize{\scriptsize}
\setlength{\tabcolsep}{1pt}
\tablecaption{Results from previous studies \label{tab:prevres}}
\tablewidth{0cm}
\tablehead{
\colhead{Reference} & \colhead{$\log$\nh} & \colhead{$\log U$} &\colhead{$\log$(\nh$U$)}    & \colhead{line} }
\startdata
\citet{matsuokaetal08} &   12. & -2.5 -- -2.0 & 9.5 -- 10 &  Ca IR Triplet \\
\citet{sigutpradhan03} & 11.6 & -2.0 & 9.6 &  their model B for \feii  \\
\citet{padovanirafanelli88} & -- & -- & 9.8$\pm 0.3$&  \hb  \\
\citet{baldwinetal96} & 12.7 & -2.5 & $\sim$10.2 & $\lambda$1900 blend 
 \enddata
\end{deluxetable}

\begin{rotate}
\begin{figure}
\includegraphics[scale=0.5]{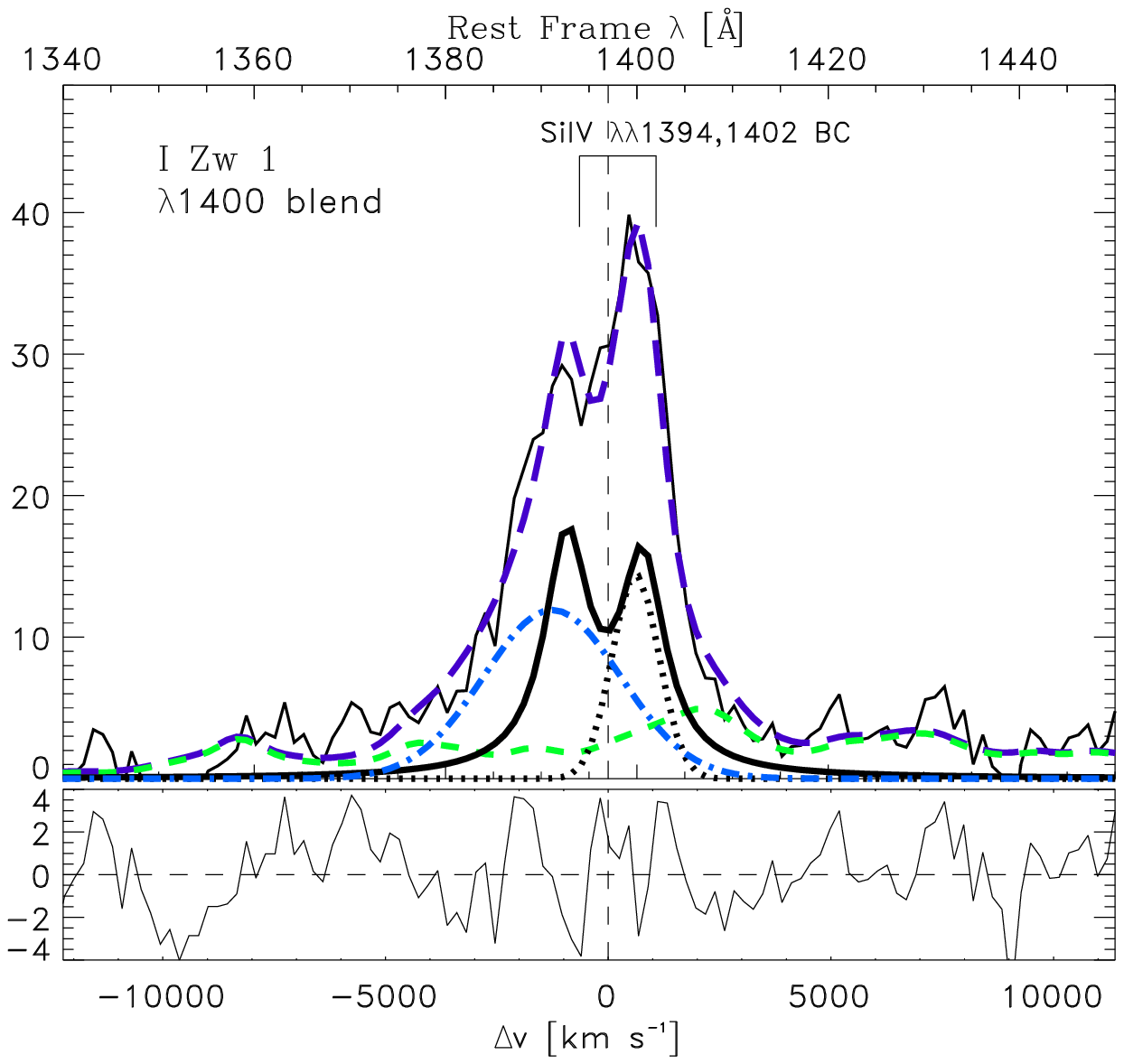}\includegraphics[scale=0.5]{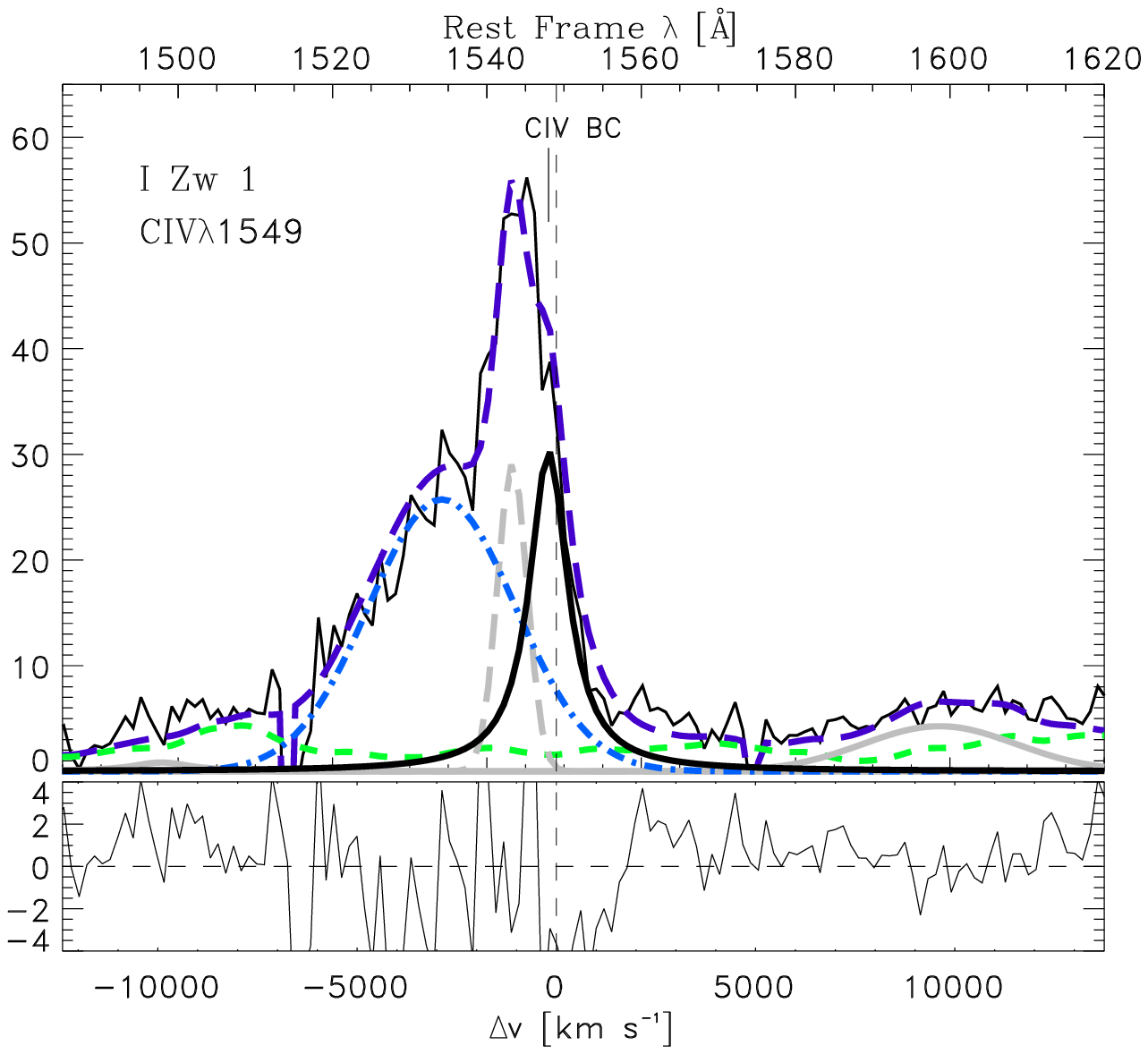}\includegraphics[scale=0.5]{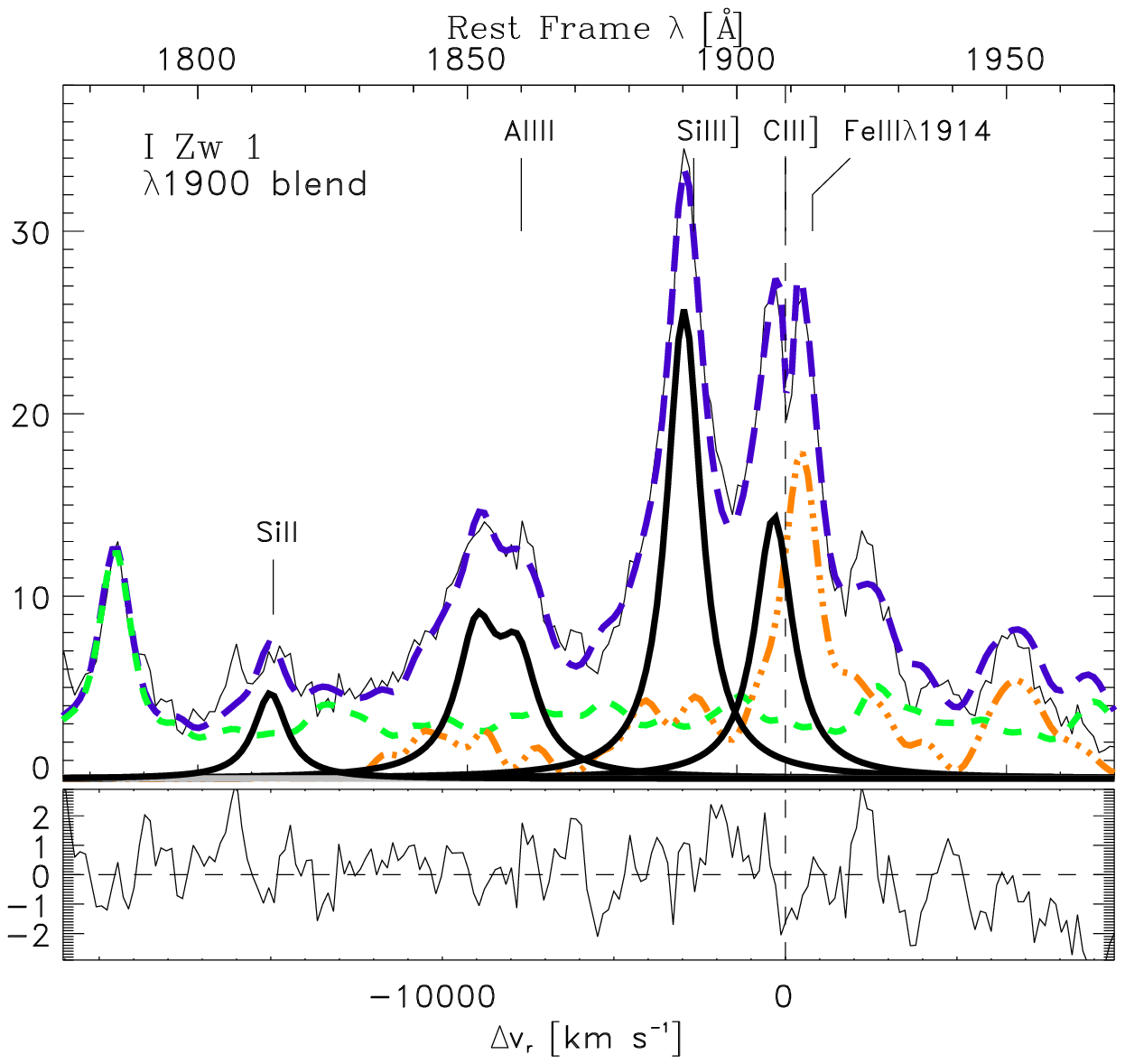}\\
\includegraphics[scale=0.5]{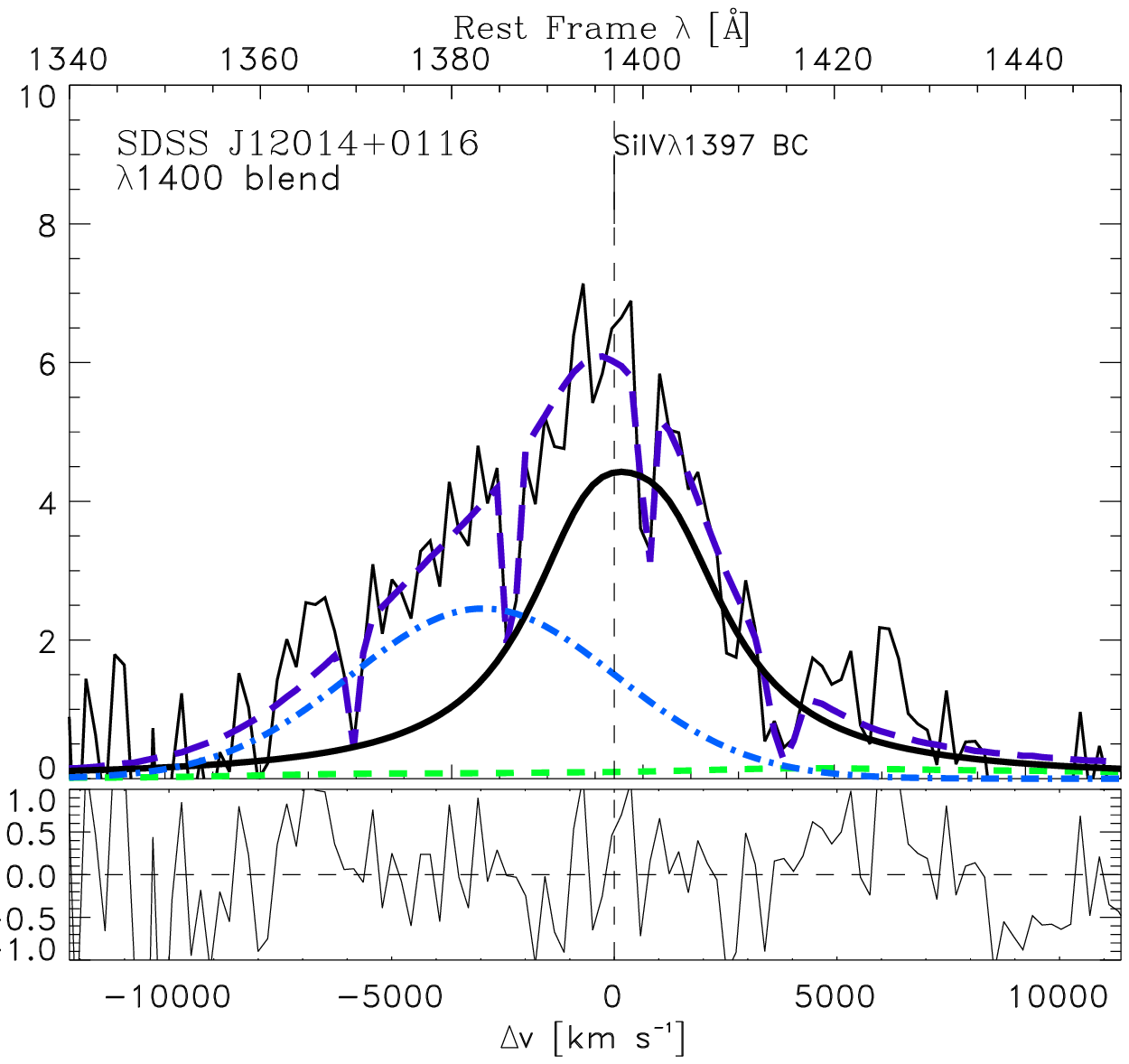}\includegraphics[scale=0.5]{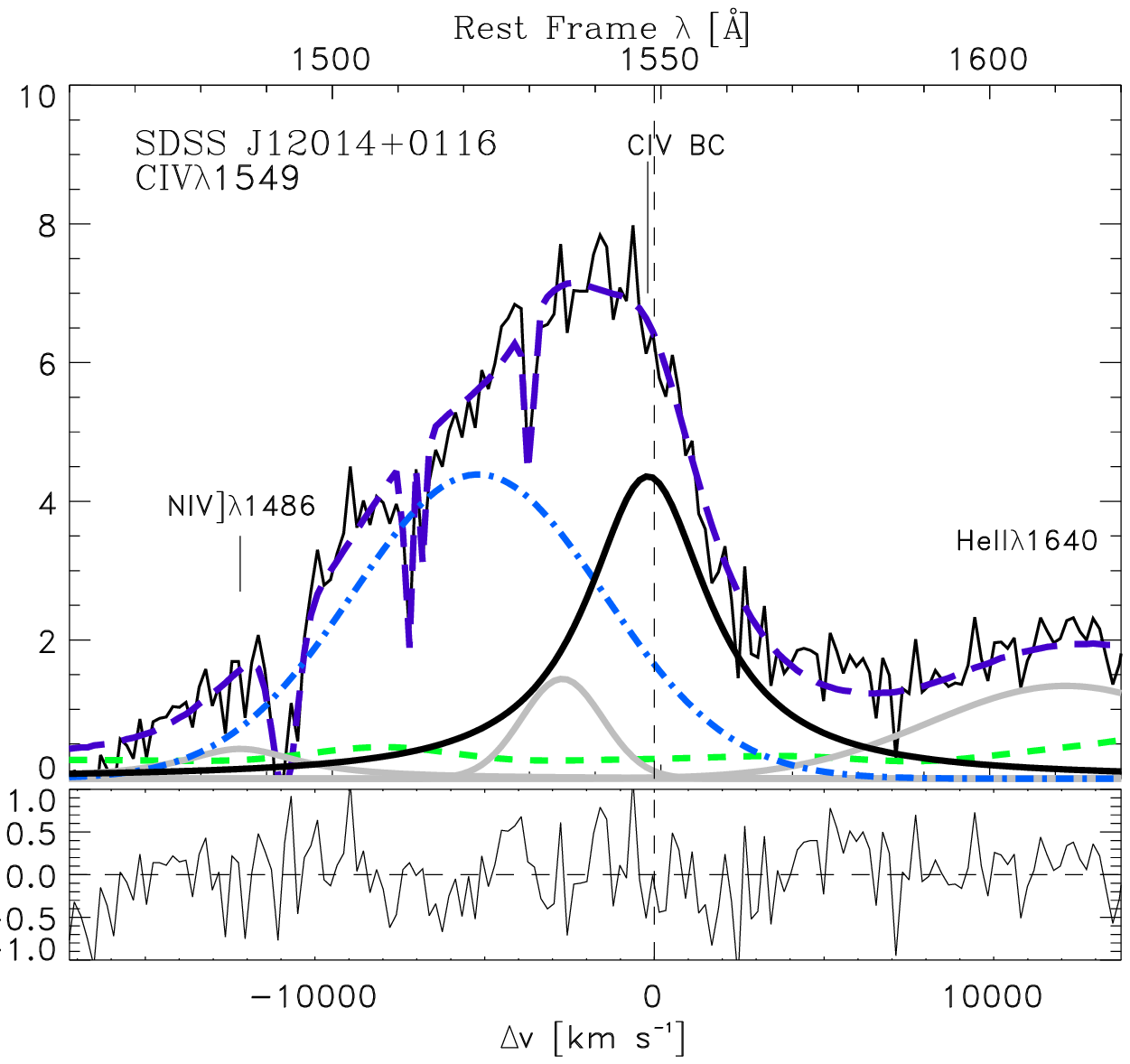}\includegraphics[scale=0.5]{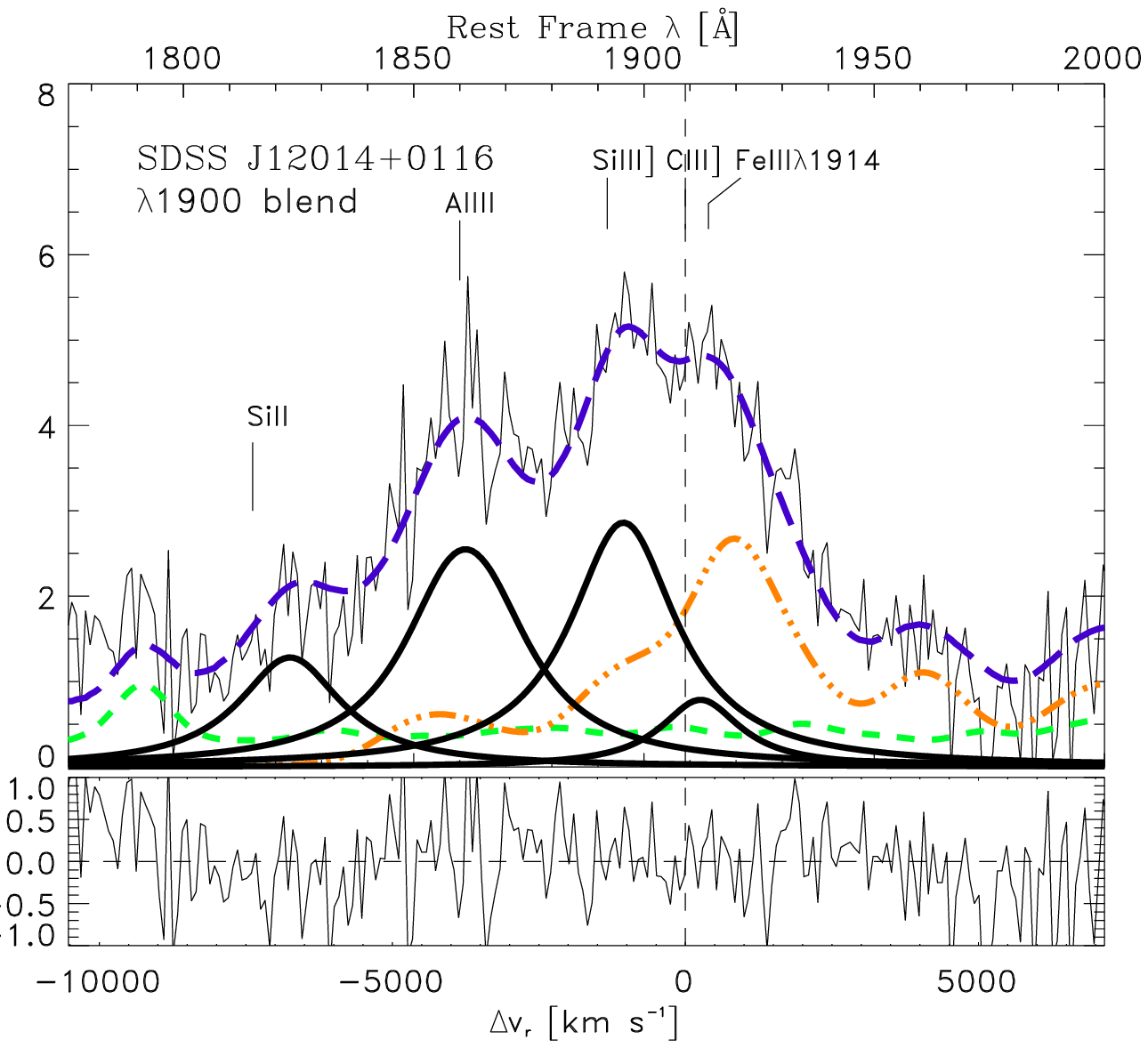}
\caption{\footnotesize We show the fits to \siiv\  (a doublet, here we show the sum, {\it left}), \civ\ ({\it middle}) and \l1900\AA\ ({\it right}) spectral region  of  I Zw 1 ({\it top}) and  SDSS J12014+0116 ({\it bottom}). The lower panels show the residuals to the fits. Upper abscissa is rest frame wavelength in \AA, lower abscissa is in velocity units. Vertical dashed line is the rest frame for \siiv, \civ\ and \ciii. The vertical scale is rest-frame specific flux in units of 10$^{-15}$\ergss\ cm$^{-2}$ \AA$^{-1}$. The long-dashed purple lines are the fits. Solid black lines are the broad central components. Short-dashed green lines represent the \feii\  template emission.
The dot-dashed blue line in \siiv\  and in \civ\ corresponds to the BLUE component. In I Zw 1, the dotted line is the contribution of \oiv, and the dashed grey line is the narrow component of \civ. For both objects the solid  grey lines  in \civ\ represents the contribution of various underlying weaker emission lines.
The major constituents of the \l1900\AA\ blend are \aliii\ (we show the sum of this doublet), \siiii\ and \feiii. The triple-dotted-dashed orange line is the \feiii\ template plus the \feiii\l1914 emission line.  See text for details. \label{fig:fits}}
\end{figure}
\end{rotate}

\begin{figure}
\includegraphics[scale=.8]{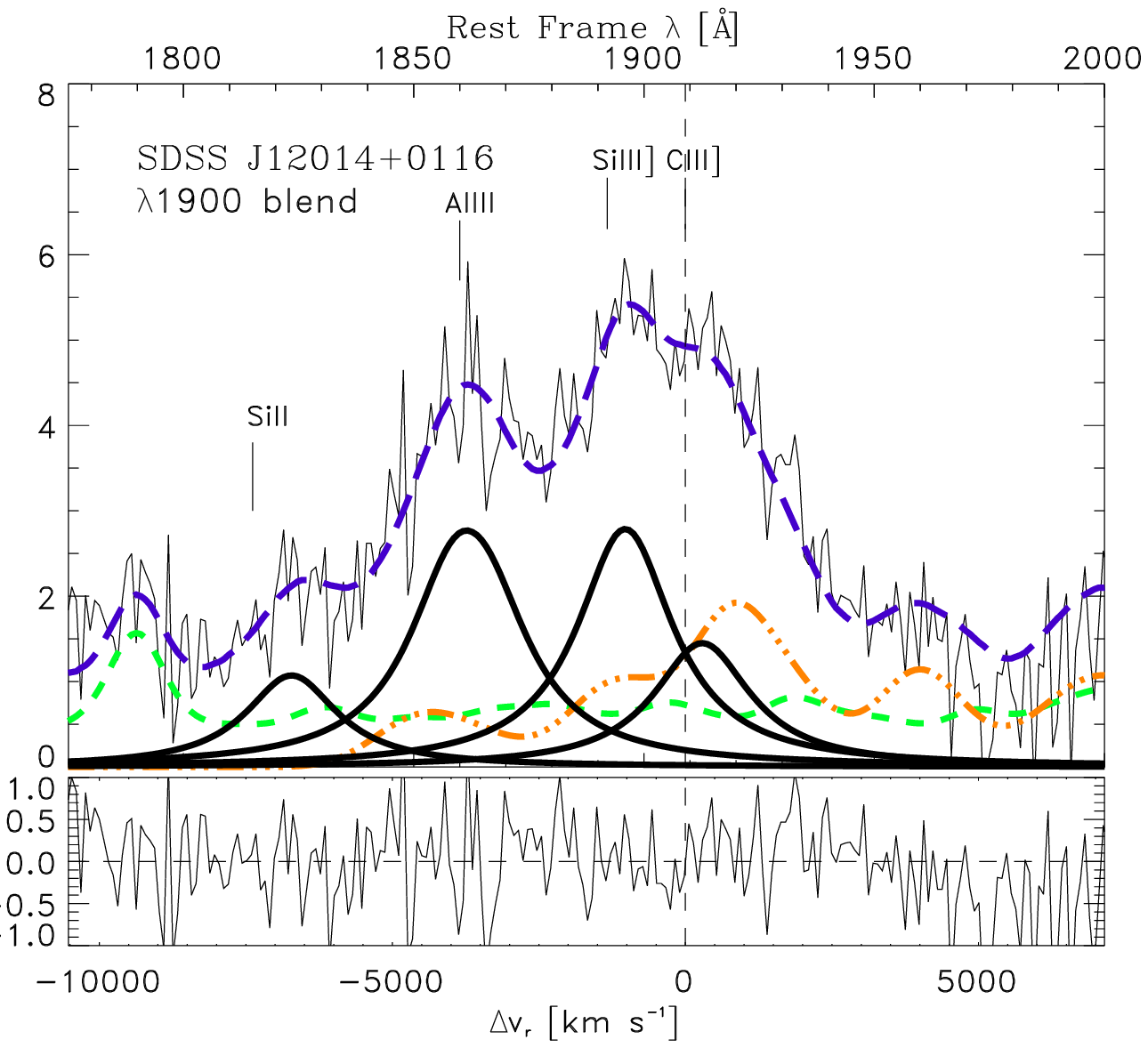}
\caption{Fits to the \l1900\AA\ blend of SDSS J12014+0116. In contrast to the lower right panel of Figure \ref{fig:fits}, here we show the maximum contribution of \ciii\ line, when the \feiii\l1914 emission line is absent. 
The triple-dotted-dashed orange line is the \feiii\ template only. 
Units and symbols are the same as in Fig. \ref{fig:fits}. See text for details. \label{fig:fitsdssciii}}
\end{figure}

\begin{figure}
\includegraphics[angle=0,scale=.4]{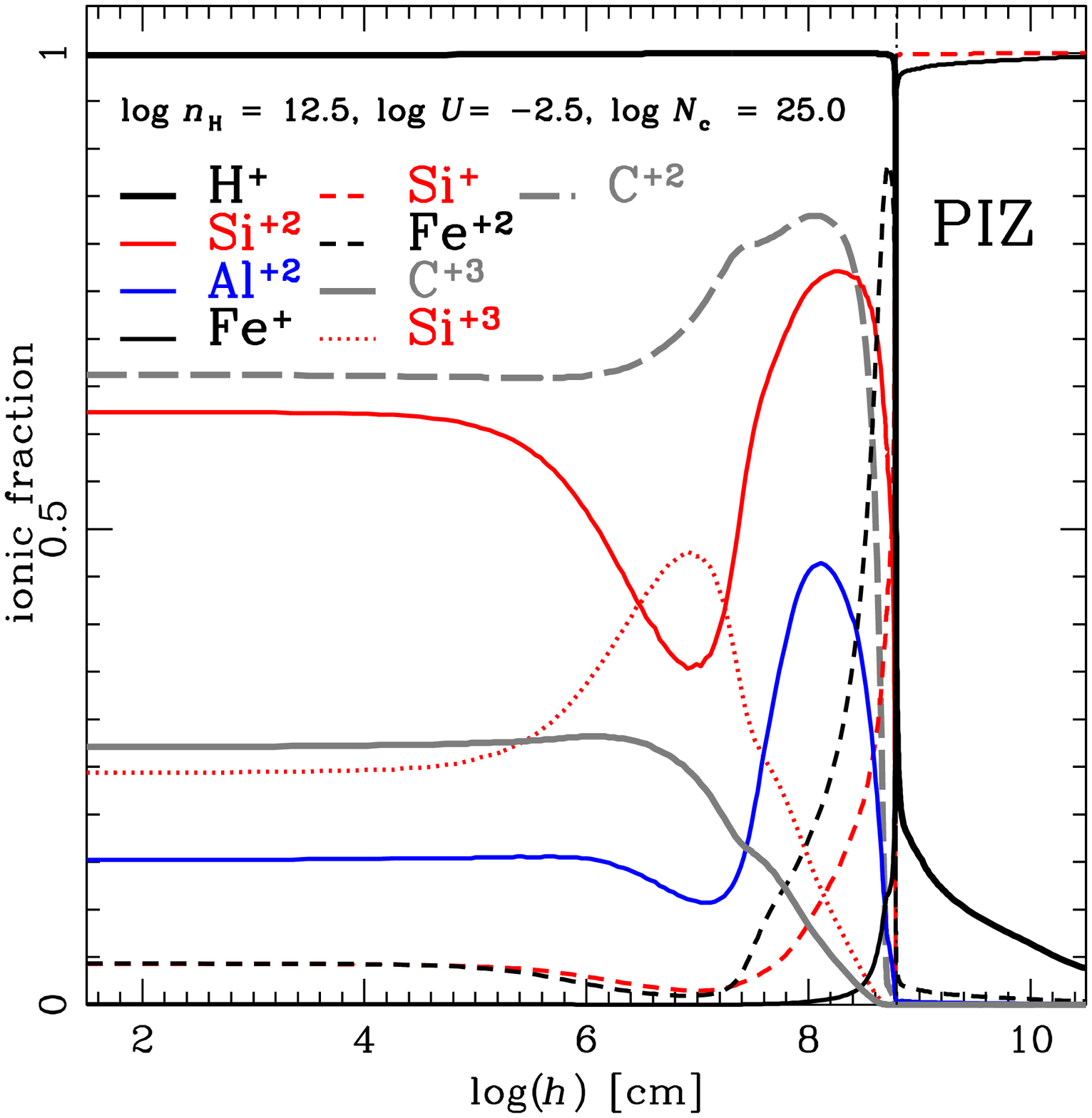}
\includegraphics[angle=0,scale=.4]{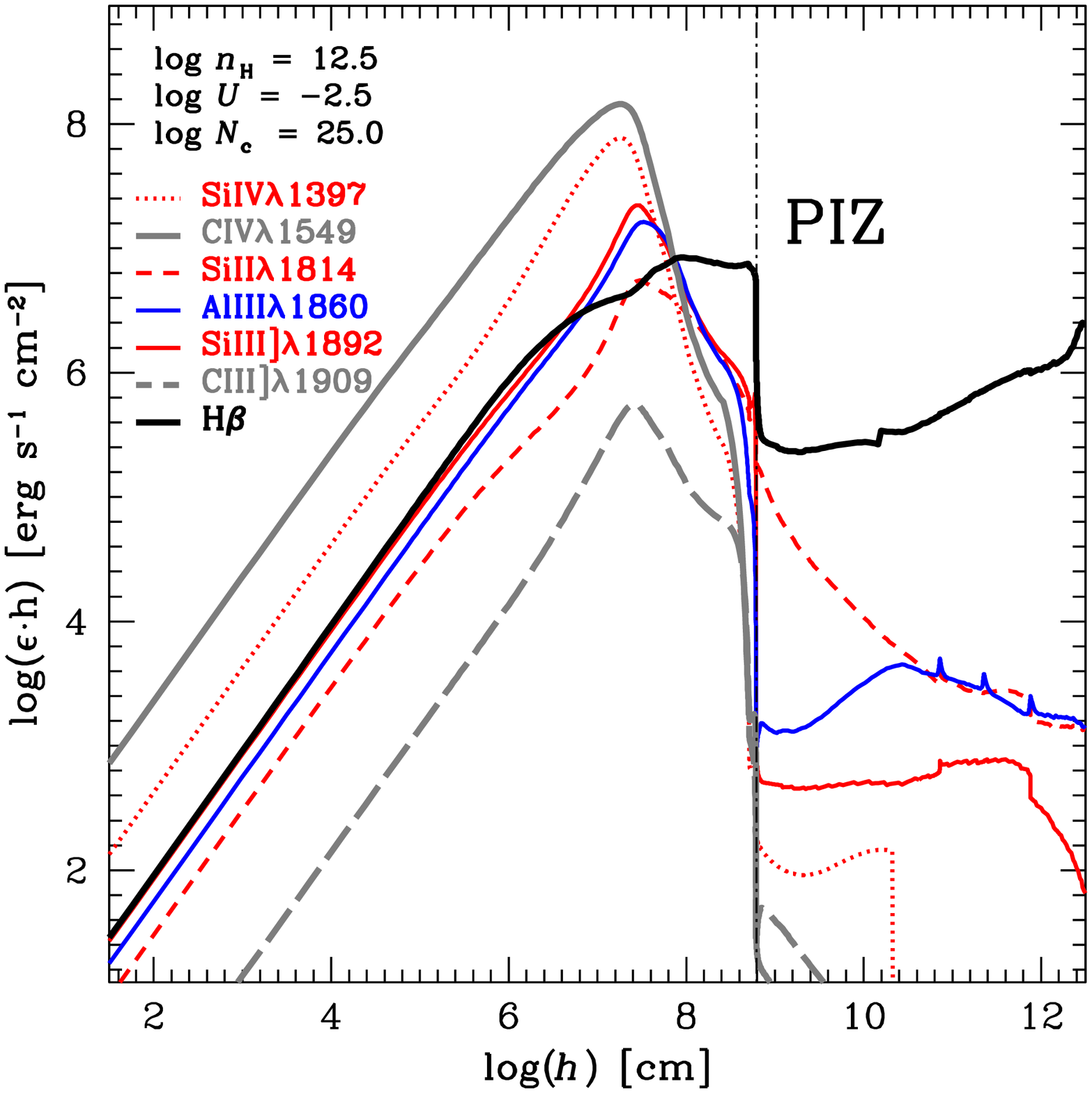}
\caption{We show the result of CLOUDY simulations for the behavior of the emission lines in a gas slab. Left: Ionic Fractions as a function of the logarithm of geometric depth $h$ in a  gas slab.   Plotted ionic stages   (H$^+$, thick solid black; Si$^+2$, solid red; Al$^+2$, solid blue; Fe$^+$,  solid black; Si$^+$, dashed red; Fe$^+2$, dashed black; C$^+3$, thick grey; Si$^+3$, dotted red) are the ones relevant to the emission lines considered in this paper.  Right: Local line emissivity per unit volume in units of \ergss\ \cm3\ multiplied by depth $h$ in cm as a function of the logarithm of $h$ for \siiv\ (red dotted), \civ\ (thick grey), \siii\ (dashed red), \aliii\ (solid blue), \siiii\ (solid red), \ciii\ (thick dashed grey), \hb\ (thick black). The continuum photons enter from left. The partially ionized zone (PIZ) is on the right side of the dot-dashed line. Ionic fraction and emissivity are calculated through a dedicated {\sc cloudy} simulation extending up to \nc = 10$^{25}$ cm$^{-2}$. \label{fig:ionic_emissivity}}
\end{figure}

\begin{figure}
\includegraphics[scale=0.7]{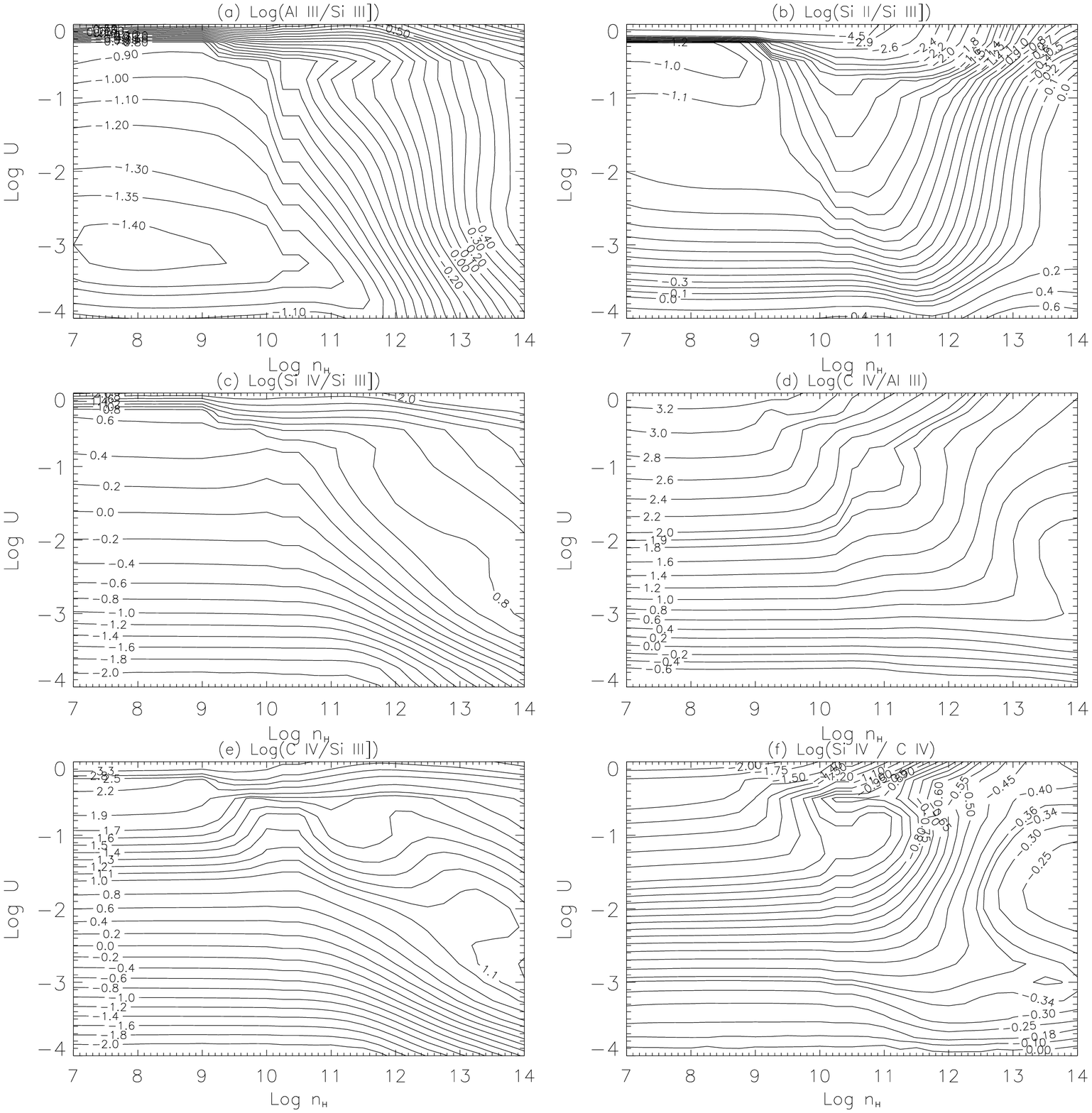}
\caption{Iso-contours of the CLOUDY simulations of the line ratios (a)  $\log$(\aliii/\siiii), (b) $\log$(\siii/\siiii), (c) $\log$(\siiv/\siiii), (d) $\log$(\civ/\aliii), (e) $\log$(\civ/\siiii) and (f) $\log$(\siiv/\civ). We use a standard value of \nc=$10^{23}$ and solar metallicity. Abscissa is hydrogen density in cm$^{-3}$, ordinate is the ionization parameter, both in logarithm scale.\label{fig:contours}}
\end{figure}

\begin{figure}
\includegraphics[scale=0.45]{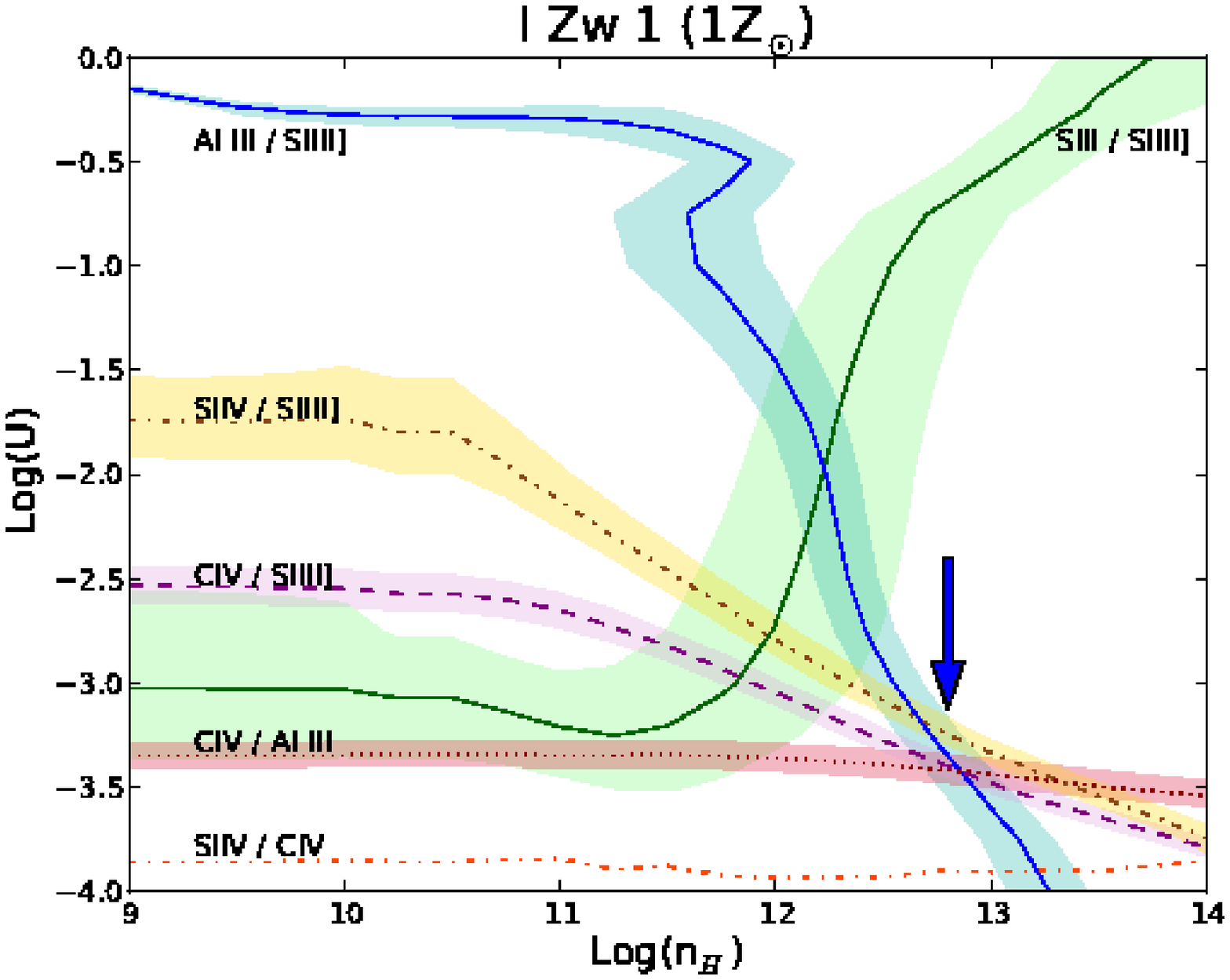}\includegraphics[scale=0.45]{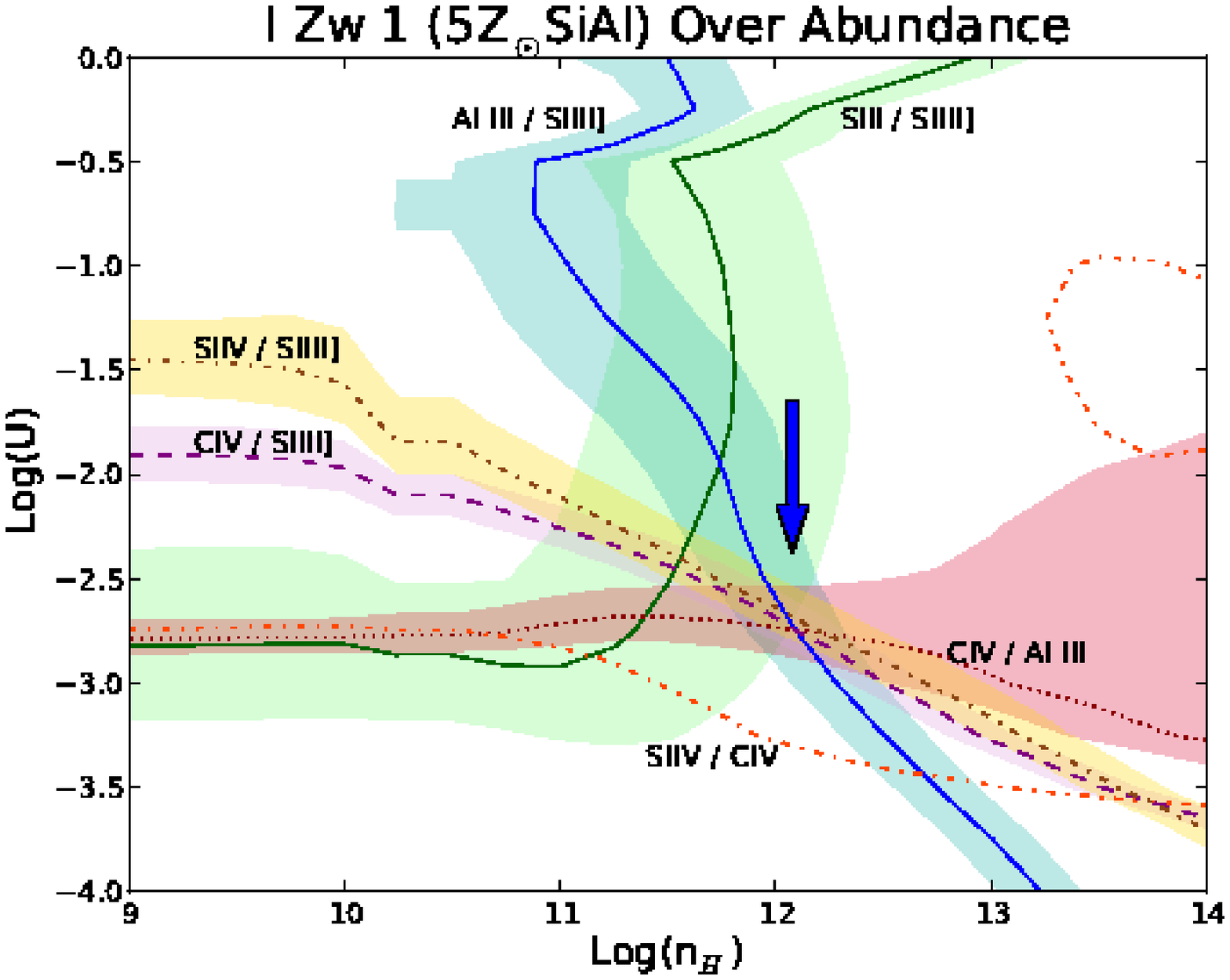}\\
\includegraphics[scale=0.45]{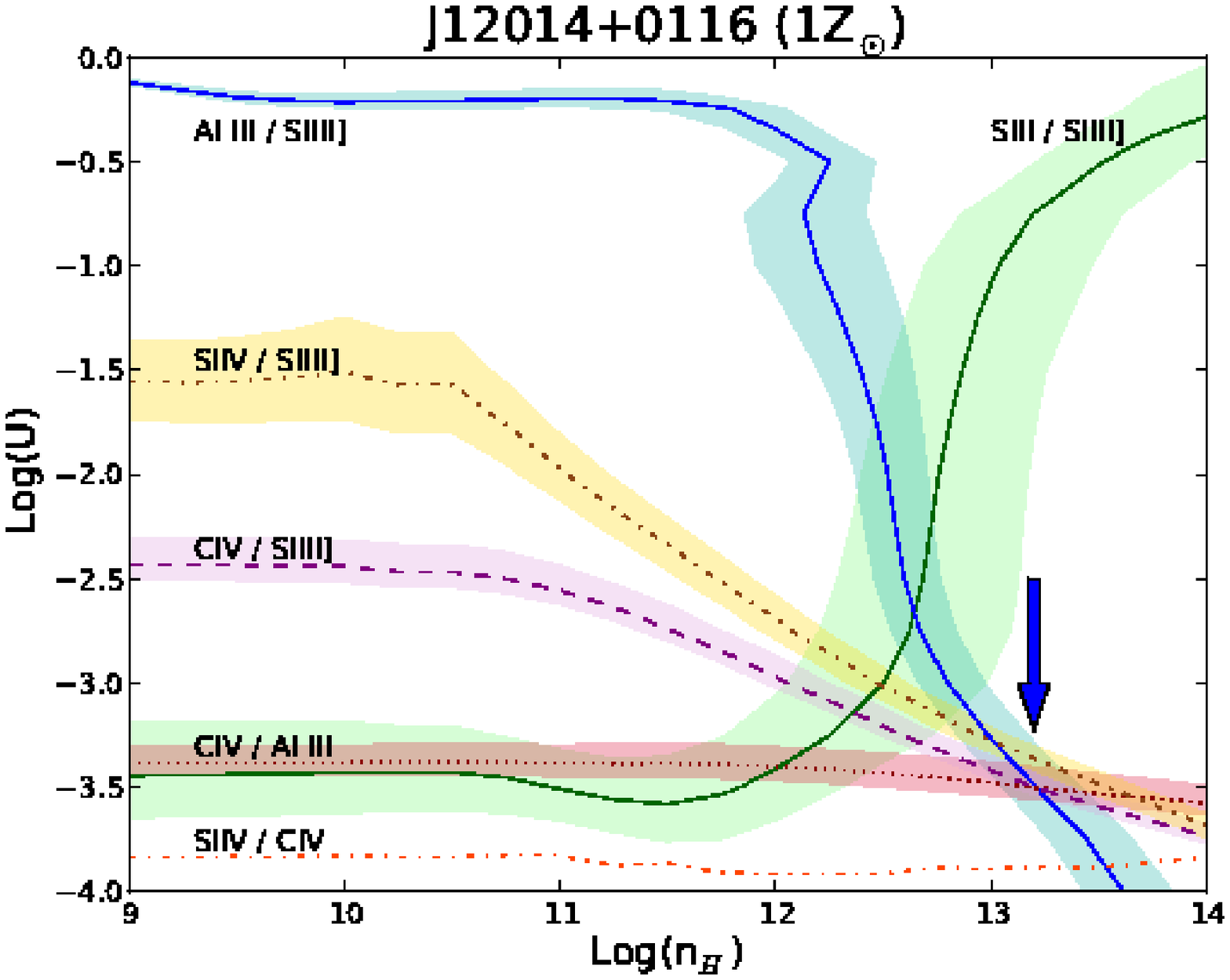}\includegraphics[scale=0.45]{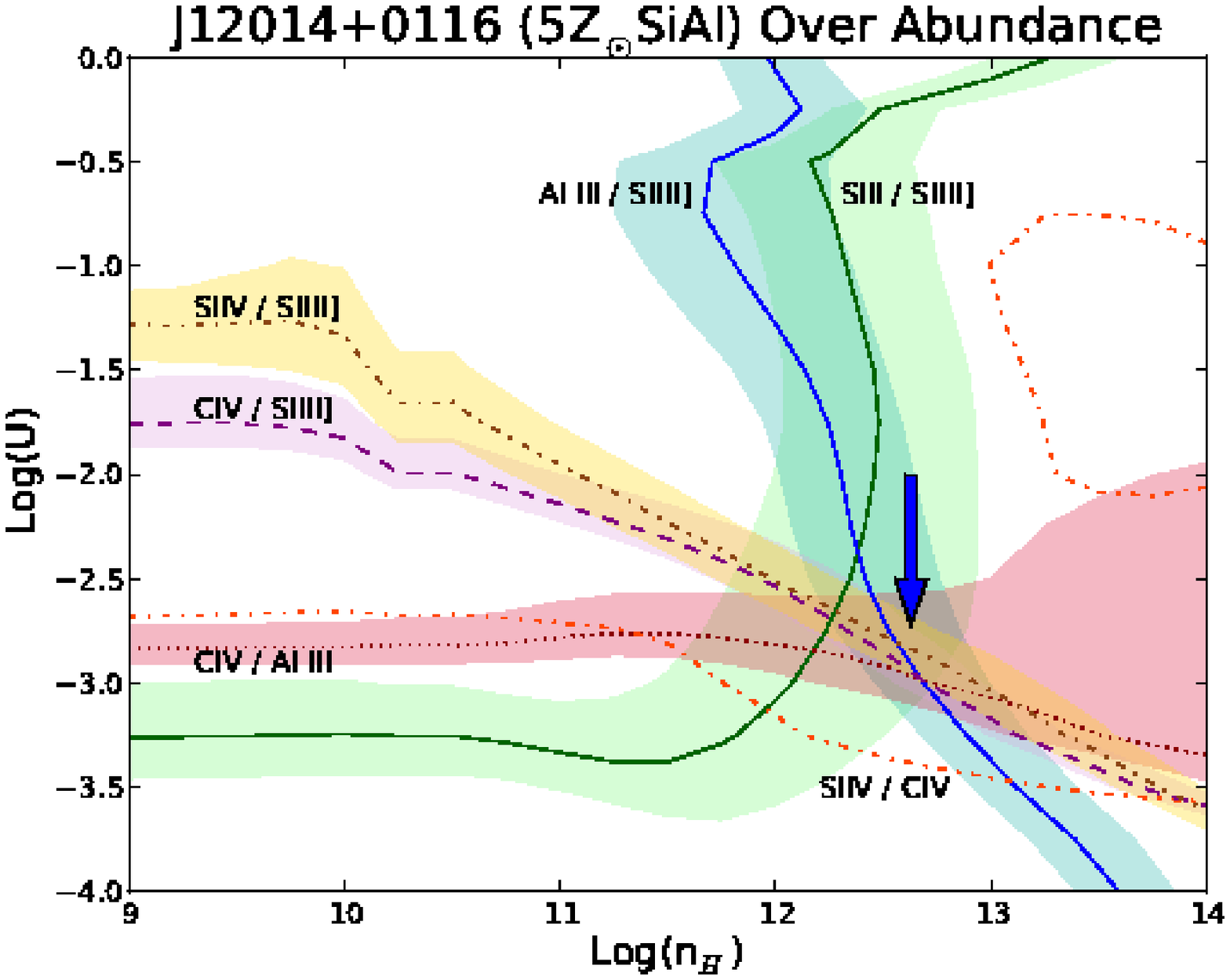}\\
\caption{Iso-contours for I Zw 1 (top) and SDSS J12014+0116 (bottom). We select the contours of the CLOUDY simulations from Figure \ref{fig:contours},  which correspond to the measured line ratio from the observed spectra (Table \ref{tab:obs}). The left panels refer to the case of solar metallicity; the right ones to the case of five times solar plus  an overabundance of Si and Al with respect to carbon by a factor 3, described in \S\ref{interp} (5Z$_{\odot}$SiAl). Each arrow points  toward the point of convergence that defines the most likely value of $U$\ and \nh. The bands are the uncertainty bands of the ratios (except for \siiv/\civ). Note that the \siiv/\civ\ ratio is not a useful constraint in the 5Z$_{\odot}$SiAl case, since it varies little in a wide area around the intersection point in the (\nh, $U$) plane.
\label{fig:contour}}
\end{figure}

\end{document}